\documentclass[preprint2]{proto}
\usepackage{times}
\usepackage{color}
\usepackage{graphicx}

\newcommand{\gsim}{\mathrel{\hbox{\rlap{\lower.55ex \hbox {$\sim$}}
                   \kern-.3em \raise.4ex \hbox{$>$}}}}
\newcommand{\lsim}{\mathrel{\hbox{\rlap{\lower.55ex \hbox {$\sim$}}
                   \kern-.3em \raise.4ex \hbox{$<$}}}}
\newcommand{\sm}[1]{\mbox{{\scriptsize #1}}}

\newcommand{\be}{\begin{equation}}
\newcommand{\ee}{\end{equation}}
\newcommand{\bea}{\begin{eqnarray}}
\newcommand{\eea}{\end{eqnarray}}
\newcommand{\bdm}{\begin{displaymath}}
\newcommand{\edm}{\end{displaymath}}
\newcommand{\bef}{\begin{figure}}
\newcommand{\eef}{\end{figure}}
\newcommand{\befone}{
  \begin{figure*}
  \centering
  \begin{minipage}{\textwidth}
  }
\newcommand{\eefone}{\end{minipage}\end{figure*}}

\newcommand{\km}{\mbox{km}}
\newcommand{\AU}{\mbox{AU}}
\renewcommand{\sec}{\mbox{s}}

\newcommand{\Lsol}{\mbox{$L_{\odot}$}}

\newcommand{\au}{\mbox{AU}}

\newcommand{\ys}{\mbox{years}}

\begin{document}

\title{\textbf{\LARGE The Earliest Stages of Star and Planet Formation:\\
Core Collapse, and the Formation of Disks and Outflows}}

\author {\textbf{\large Zhi-Yun Li}}
\affil{\small\em University of Virginia}

\author {\textbf{\large Robi Banerjee}}
\affil{\small\em Universit{\"a}t Hamburg}

\author {\textbf{\large Ralph E.~Pudritz}}
\affil{\small\em McMaster University}

\author {\textbf{\large Jes K.~J{\o}rgensen}}
\affil{\small\em Copenhagen University}

\author {\textbf{\large Hsien Shang and Ruben Krasnopolsky}}
\affil{\small\em Academia Sinica}

\author {\textbf{\large Ana{\"e}lle Maury}}
\affil{\small\em Harvard-Smithsonian Center for Astrophysics}

\begin{abstract}
\baselineskip = 11pt
\leftskip = 0.65in
\rightskip = 0.65in
\parindent=1pc
{\small

The formation of stars and planets are connected through disks. Our
theoretical understanding of disk formation has undergone drastic
changes in recent years, and we are on the brink of a
revolution in disk observation enabled by ALMA\@.  Large rotationally supported
circumstellar disks, although
common around more evolved young stellar objects, are rarely
detected during the earliest, ``Class 0'' phase; a few excellent
candidates have been discovered recently around both low- and
high-mass protostars though.
In this early phase,
prominent outflows are ubiquitously observed; they are expected to be
associated with at least small magnetized disks.  Whether the paucity
of large Keplerian disks is due to observational challenges
or intrinsically
different properties of the youngest disks is unclear. In this review
we focus on the observations and theory of the formation of early
disks and outflows, and their connections with the first phases of
planet formation. Disk formation --- once thought to be a simple
consequence of the conservation of angular momentum during hydrodynamic core
collapse --- is far
more subtle in magnetized gas. In this case, the rotation can be
strongly magnetically braked. Indeed, both analytic arguments and
numerical simulations have shown that disk formation is suppressed in
the strict ideal magnetohydrodynamic (MHD) limit for the observed level of core
magnetization.
We review what is known about this ``magnetic braking
catastrophe'', possible ways to resolve it, and the current status of
early disk observations. Possible resolutions include non-ideal MHD effects
(ambipolar diffusion, Ohmic dissipation and Hall effect), magnetic
interchange instability in the inner part of protostellar accretion
flow, turbulence, misalignment between the magnetic field and rotation
axis, and depletion of the slowly rotating envelope by outflow
stripping or accretion.
Outflows are also intimately linked to disk formation; they
are a natural product of magnetic fields and rotation and are
important signposts of star formation. We review new developments
on early outflow generation since PPV\@. The properties of
early disks and outflows are a key component of planet formation in
its early stages and we review these major connections.
\\~\\~\\~}

\end{abstract}

\section{Overview}

This review focuses on the earliest stages of star and
planet formation, with an emphasis on the origins of early disks
and outflows, and conditions that characterize the earliest stages
of planet formation.

The importance of disks is obvious. They are the birthplace of
planets, including those in our solar system. Nearly
1000 exoplanets have been discovered to date (http://exoplanet.eu;
see the chapters by Chabrier et al.\ and Helled et al.\ in this volume).
The prevalence of planets indicates that disks must be common at least at
some point in time around Sun-like stars. Observations show that this
is indeed the case.

Direct evidence for circumstellar disks around young stellar objects
first came from the HST observations of the so-called Orion
``proplyds'' \citep{OdellWen92,MccaughreanOdell96}, where
the disks are seen in silhouette against the bright background. More
recently, with the advent of millimeter and sub-millimeter arrays,
there is now clear evidence from molecular line observations that
some protoplanetary disks have Keplerian velocity fields, indicating
rotational support (e.g., ALMA observations of TW Hydra; see
\S~\ref{Observations} on observations of early disks).
Indirect evidence, such as protostellar outflows and infrared
excess in spectral energy distribution, indicates that
the majority, if not all, low-mass, Sun-like stars pass through
a stage with disks, in agreement with the common occurrence of
exoplanets.

Theoretically, disk formation --- once thought to be a trivial
consequence of the conservation of angular momentum during
hydrodynamic core collapse --- is far more subtle in magnetized
gas. In the latter case, the rotation can be strongly magnetically
braked. Indeed, disk formation is suppressed in the strict ideal
MHD limit for the observed level of core magnetization;
the angular momentum of the
idealized collapsing core is nearly completely removed by magnetic
braking close to the central object.  How is this resolved?

We review what is known about this so-called ``magnetic braking
catastrophe'' and its possible resolutions (\S~\ref{DiskTheory}).
Important processes to be discussed include non-ideal MHD effects
(ambipolar diffusion, Ohmic dissipation and Hall effect),
magnetic interchange instability in the
inner part of protostellar accretion flow, turbulence, misalignment
between the magnetic field and rotation axis, and depletion of the
slowly rotating envelope by outflow stripping or accretion.
We then turn to a discussion of the launch of the earliest outflows,
and show that two aspects of such outflows ---
the outer magnetic ``tower'' and the inner centrifugally driven disk
wind that have dominated much of the discussion and theory of early
outflows, are actually two regimes of the same unified MHD theory
(\S~\ref{outflow}). In \S~\ref{Disk2Planet}, we discuss the earliest
phases of planet formation in such disks which are likely quite
massive and affected by the angular momentum transport via both
strong spiral waves and powerful outflows. We synthesize the results
in \S~\ref{Synthesis}.

\section{Observations}
\label{Observations}

\subsection{Dense Cores}
\label{ObsCores}

Dense cores are the basic units for the formation of at least low-mass
stars. Their properties determine the characteristics of the disk,
outflow and planets---the byproducts of the star formation process.
Particularly important for the formation of the outflow and disk (and
its embedded planets) are the core rotation rate and magnetic field
strength. We begin our discussion of observations with these two key
quantities, before moving on to early outflows and disks.

Rotation rate is typically inferred from the velocity gradient
measured across a dense core \citep{Goodman93,Caselli02}.
Whether the gradient traces true rotation or not is
still under debate \citep{Bergin07,Dib10}.
For example, infall along a filament can mimic
rotation signature (\citealt{Tobin12_gradients}; for synthetic line
emission maps from filament accretion and their interpretation, see
also \citealt{Smith12,Smith13}).
If the gradient does trace
rotation, then the rotational energy of the core would typically be
a few percent of the gravitational energy. Such a rotation would not
be fast enough to prevent the dense core from gravitational collapse.
It is, however, more than enough to form a large, $10^2$~AU-sized
circumstellar disk, if angular momentum is conserved during the core
collapse.

Magnetic fields are observed in the interstellar medium on a wide
range of scales (see chapter by H.-B. Li et al.\ in this volume).
Their dynamical importance relative to gravity is usually
  measured by the ratio of the mass of a region to the magnetic flux
  threading the region. On the core scale, the field strength
was characterized
by \citet{Troland08}, who carried out an OH Zeeman survey
of a sample of dense cores in nearby dark clouds. They inferred a
median value $\lambda_{los}\approx 4.8$ for the dimensionless
mass-to-flux ratio (in units of the critical value $[2\pi
G^{1/2}]^{-1}$, \citealt{Nakano78}). Geometric corrections should
reduce the ratio to a typical value of $\lambda \approx 2$
\citep{Crutcher12}.
It corresponds to a ratio of magnetic to gravitational energy of
tens of percent, since the ratio is given roughly by $\lambda^{-2}$.
Such a field is not strong enough to prevent the core from collapsing
into stars. It is, however, strong enough to dominate the rotation in
terms of energy and, therefore, is expected to strongly affect disk
formation (\S~\ref{DiskTheory}).

\subsection{Early Outflows}
\label{ObsOutflows}

Jets and outflows are observed during the formation of stars over the
whole stellar spectrum, from brown dwarfs \citep[e.g.,][]{Whelan05,
Whelan12} to high-mass stars \citep[e.g.,][]{Motogi13, Carrasco10,
Qiu11, Qiu08a, Qiu07a, Zhang07}, strongly indicating a universal
launching mechanism at work (see \S~\ref{outflow} and also chapter
by Frank et al.\ in this volume).

Young brown dwarfs have optical forbidden line spectra similar to
those of low-mass young stars that are indicative of outflows
\citep[][]{Whelan05,Whelan06}. The inferred outflow speeds
of order $40$--$80\,\km\,\sec^{-1}$ are somewhat lower than those of
young stars \citep{Whelan07, Joergens13}. \citet{Whelan12} suggest
that brown dwarf outflows can be collimated and episodic, just
as their low-mass star counterparts. There is some indication
that the ratio of outflow and accretion rates, $\dot{M}_{\sm{out}}
/\dot{M}_{\sm{accr}}$, is higher for young brown dwarf
and very-low mass stars \citep{Comeron03, Whelan09, Bacciotti11}
than for the classical T-Tauri stars \citep[e.g.,][]{Hartigan95,
Sicilia-Aguilar10, Fang09, Ray07}. Whether this is generally true
remains to be established.

Since planetary systems like the Jovian system with its Galilean moons
are thought to be built up from planetary sub-disks \citep{Mohanty07},
one would naturally expect them to launch outflows as well
\citep[see e.g.,][]{Machida06, Liu09}. There is, however,
no direct observational evidence yet for such circum-planetary disk-driven
outflows.

At the other end of the mass spectrum, there is now evidence that
outflows around young massive stars can be highly collimated, even
at relative late evolutionary stages \citep[e.g.,][]{Carrasco10,
Rodriguez12, Chibueze12, Palau13}. For example, interferometric
observations reveal that the young, luminous (~$\sim 10^5 \Lsol$),
object IRAS 19520+2759 drives a well collimated CO outflow, with
a collimation factor of 5.7 \citep{Palau13}. It appears to have
evolved beyond a central B-type object, but still drives a
collimated outflow, in contrast with the expectation that massive
YSO outflows decollimate as they evolve in time \citep{Beuther05}.
Interestingly, an HII region has yet to develop in this source.
It may be quenched by protostellar accretion flow \citep{Keto02,
Keto03, Peters10a, Peters11a} or absent because the central stellar
object is puffed up by rapid accretion (and thus not hot enough at
surface to produce ionizing radiation; \citealt{Hosokawa10}).

\subsection{Early Disks}
\label{ObsDisks}
From an observational point of view, the key question to address is
when rotationally supported
circumstellar disks are first established and become observable.
It is clear that, after approximately 0.5~Myr \citep{evans09},
gaseous Keplerian (protoplanetary) disks are present on the
scales of $\sim$100--500~AU around both low- and intermediate-mass
stars (``T Tauri'' and ``Herbig Ae'' stars, respectively --- or
Class~II young stellar objects; see \citealt{dutrey07}). Whether they
are present at earlier times requires high-resolution studies of
the youngest protostars, e.g., Class~0 and Class~I objects.

\begin{figure}
  \resizebox{\hsize}{!}{\includegraphics{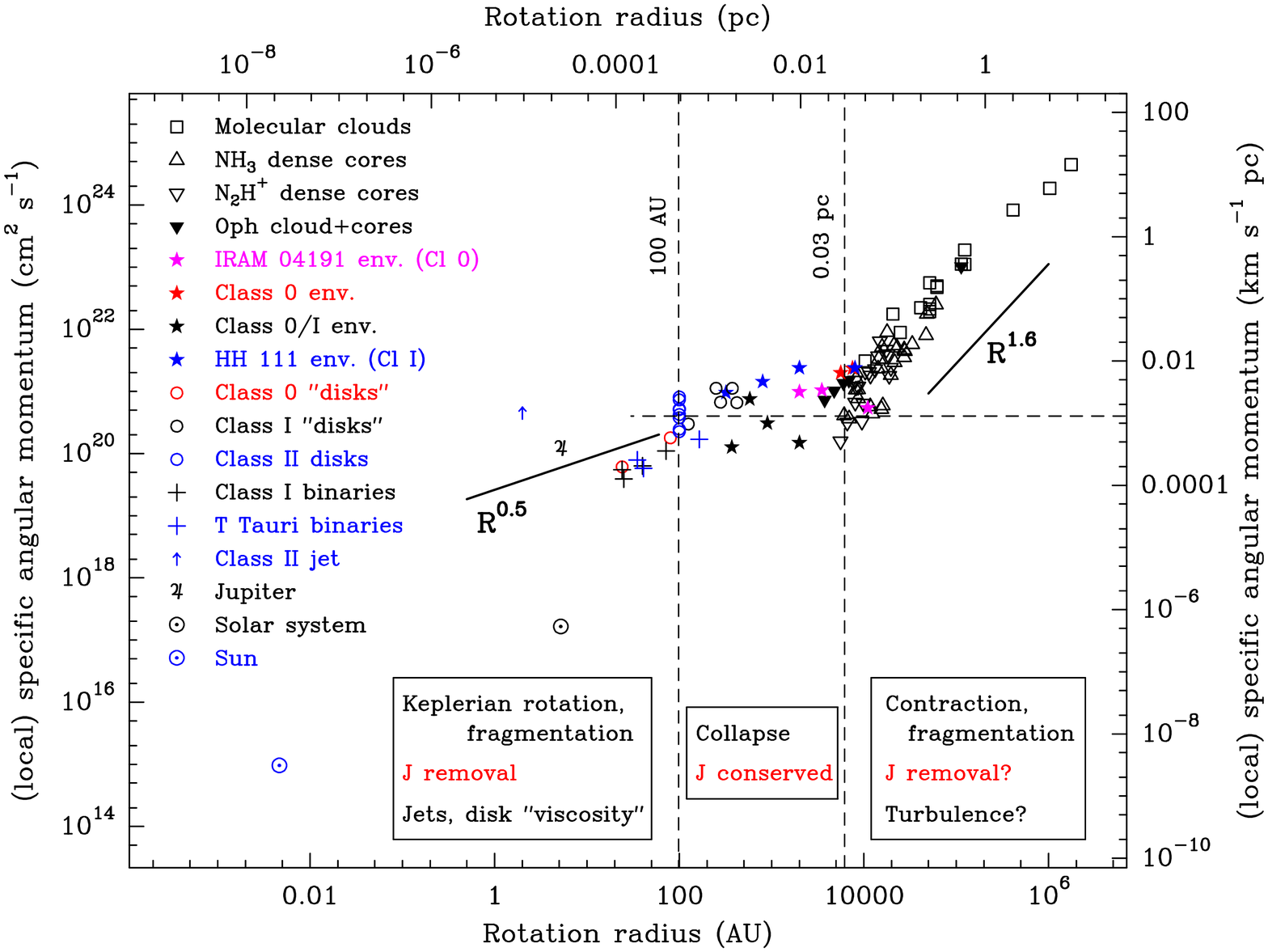}}
  \caption{Progression of specific angular momentum as function of
    scale and/or evolutionary stage of young stellar objects. Figure
    from \cite{Belloche13}.}\label{angularmomentumfigs}
\end{figure}

One way to constrain the process of disk formation is to study the
rotation rates on difference scales (see the caveats
in inferring rotation rate in \S~\ref{ObsCores}). From larger to
smaller scales, a clear progression in kinematics is evident
(Fig.~\ref{angularmomentumfigs}; reproduced from
\citealt{Belloche13}): at large distances from the central protostar
and in prestellar cores, the specific angular momentum decreases
rapidly toward smaller radii, implying that the angular velocity
is roughly constant
\citep[e.g.,][]{Goodman93,Belloche02}.  Observations of objects in
relatively late stages of evolution suggest that the specific
angular momentum tends to a constant value ($v_{\rm rot} \propto
r^{-1}$)
between $\sim 10^2$ and $10^4$~AU, as expected from
conservation of angular momentum under infall. A rotationally
supported disk is expected to show increasing specific angular
momentum as function of radius (Keplerian rotation, $v_{\rm rot}
\propto r^{-0.5}$). To characterize the properties of disks being
formed, the task at hand is to search for the location where the
latter two regimes, Keplerian disk with $v_{\rm rot} \propto r^{-0.5}$
and the infalling envelope with $v_{\rm rot}\propto r^{-1}$, separate.

\subsubsection{Techniques}
\label{techniques}

Observationally the main challenge in revealing the earliest stages of
the circumstellar disks is the presence of the larger scale
protostellar envelopes during the Class~0 and I stages, which
reprocess most of the emission from the central protostellar object
itself at shorter wavelengths and easily dominate the total flux at
longer wavelengths. The key observational tools at different
wavelength regimes are (for convenience we define the
  near-infrared as $\lambda$$<$3~$\mu$m, mid-infrared as
  3~$\mu$m$<$$\lambda$$<$50~$\mu$m, far-infrared as
  50~$\mu$m$<$$\lambda$$<$250~$\mu$m, (sub)millimeter as
  250~$\mu$m$<$$\lambda$$<$4~mm):

\begin{description}
\item[Mid-infrared:] at mid-infrared wavelengths the observational
  signatures of young stars are dominated by the balance between the
  presence of warm dust and degree of extinction on small-scales ---
  and in particular, highly sensitive observations with the
  \emph{Spitzer Space Telescope} in 2003--2009 have been instrumental
  in this field. One of the key results relevant to disk formation
  during the embedded stages is that simple infalling envelope
  profiles (e.g.\ $\rho \propto r^{-2}$ or $r^{-1.5}$) cannot extend
  unmodified to $\lesssim$~few hundred AU scales
  \citep{iras16293letter,enoch09}: the embedded protostars are
  brighter in the mid-infrared than expected from such profiles, which
  can for example be explained if the envelope is flattened or
  has a cavity on small scales. Indeed, extinction maps reveal
  that protostellar environments are complex on $10^3$~AU scales
  \citep{Tobin10} and in some cases show asymmetric and filamentary
  structures that complicate the canonical picture of formation of
  stars from the collapse of relatively spherical dense cores.

\item[Far-infrared:] at far-infrared wavelengths the continuum
  emission comes mainly from thermal dust grains with temperatures of
  a few tens of K\@. This wavelength range is accessible almost
exclusively from space only. Consequently, observatories such as
the Infrared Space
  Observatory (ISO) and Herschel Space Observatory \citep{Pilbratt10}
  have been the main tools for characterizing
  protostars there. At the time of writing, surveys from the Herschel
  Space Observatory are starting to produce large samples of deeply
  embedded protostellar cores that can be followed up by other
  instruments. The observations at far-infrared wavelengths
  provide important information about the peak of the luminosity of
  the embedded protostars and the distribution of low surface
  brightness dust --- but due to the limitation in angular resolution
  less information on the few hundred AU scales of disks around more
  evolved YSOs.

\item[(Sub)millimeter:] the (sub)millimeter wavelengths provide a
  unique window on the thermal radiation from
  the cooler dust grains on small scales \citep{Andre93,
  Chandler00}. Aperture synthesis observations at these
  wavelengths resolve scales down to $\sim$100~AU or better in nearby
  star forming regions. The flux from the thermal dust continuum
  emission is strongly increasing with frequency $\nu$ as $F_\nu \propto \nu^2$ or steeper,
  making these wavelengths ideally suited for detecting dense structures
  while discriminating from possible free-free emission. Likewise the
  high spectral resolution for a wide range of molecular rotational
  transitions can be tailored to study the structure (e.g.,
  temperature and kinematics) of the different components in the
  protostellar systems \citep{iras2sma}.
\end{description}

\subsubsection{Millimeter Continuum Surveys}
\label{millimetercontinuum}

The main observational tool for understanding disk formation is
millimeter surveys using aperture synthesis technique that probe how the matter is distributed
on the few hundred AU scales. The use of such a technique
to address this question goes
back to \cite{keene90} and \cite{terebey93} with the first larger
arcsecond scale surveys appearing in the early 2000s
\citep[e.g.,][]{looney00} and detailed radiative transfer modeling
appearing about the same time
\citep[e.g.][]{hogerheijde00sandell,harvey03,looney03,n1333i2art,iras2sma}.

The general conclusion from these studies is that, in most cases,
both the large- and small-scale continuum emission cannot be
reproduced by a single analytical model of a simple, axisymmetric
envelope. In some cases, e.g., \cite{brown00,iras2sma,enoch09}, these
structures are well resolved on a few hundred AU scales. Some
noteworthy counter-examples where no additional dust continuum
components are required include L483 \citep{l483art}, L723
\citep{girart09} and three of the nine Class 0 protostars in Serpens
surveyed by \cite{enoch11}. Typically the masses for individual
objects derived in these studies agree well with each other once
similar dust opacities and temperatures are adopted. \cite{evolpaper}
compared the dust components for a sample of 18 embedded Class 0 and I
protostars and did not find an increase in mass of the modeled compact
component with bolometric temperature as one might expect from the
growth of Keplerian disks.  \cite{evolpaper} suggested that this could
reflect the presence of the rapid formation of disk-like structure
around the most deeply embedded protostars, although the exact
kinematics of those around the most deeply embedded (Class~0) sources
were unclear. An unbiased survey of embedded protostars in Serpens
with envelope masses larger than 0.25~$M_\odot$ and luminosities
larger than 0.05~$L_\odot$ by \cite{enoch11} finds similar masses for
the compact structures around the sources in that sample. Generally
these masses are found to be small relative to the larger scale
envelopes in the Class~0 stage on 10,000~AU scales --- but still
typically one to two orders of magnitude larger than the mass on
similar, few hundred AU, scales extrapolated from the envelope.

Still, the continuum observations do not provide an unambiguous answer
to what these compact components represent. By the Class~I stage
some become the Keplerian disks surrounding the protostars. The
compact components around the Class~0 protostars could be the
precursors to these Keplerian disks. However, it is unlikely that such
massive rotationally supported disks could be stably supported given
the expected low stellar mass for the Class 0 protostars: they should
be prone to fragmentation (see \S~\ref{Disk2Planet}).

An alternative explanation for the compact dust emission detected in
interferometric continuum observations may be the presence of
``pseudo-disks''. In the presence of magnetic fields, torsional Alfv{\'e}n
waves in twisted field lines carry away angular momentum, preventing
the otherwise natural formation of large rotationally supported
disks. However, strong magnetic pinching forces deflect infalling gas
toward the equatorial plane to form a flattened structure---the
``pseudo-disk'' \citep{Galli93,Allen03b,Fromang06}. Unlike Keplerian
disks observed at later stages, this flattened inner
envelope is not supported by rotational motions, but can be partially
supported by magnetic fields.  Observationally disentangling disks and
pseudo-disks is of paramount importance since accretion onto the
protostar proceeds very differently through a rotationally-supported
disk or a magnetically-induced pseudo-disk.

\cite{Maury10} compared the results from an IRAM Plateau de Bure study
of five Class~0 to synthetic model images from three numerical
simulations --- in particular focusing on binarity and structure down
to scales of $\sim$50~AU\@. The comparison shows that magnetized models
of protostar formation including pseudo-disks agree better with the
observations than, e.g., pure hydrodynamical simulation in the case of
no initial perturbation or turbulence. With turbulence, magnetized
models can produce small disks ($\sim 100\, \au$) \citep{Seifried13},
which might still be compatible with \citet{Maury10}'s
observations. The compact continuum components could also represent
the ``magnetic walls'' modeled by \cite{Tassis05}, although in some
sources excess unresolved emission remains unaccounted for in this
model \citep{Chiang08}.

\subsubsection{Kinematics}
A number of more evolved Class~I young stellar objects show velocity
gradients that are well fitted by Keplerian profile
\citep[e.g.,][]{brinch07,lommen08,evolpaper}
(see also \citealt{Harsono13}, and \citealt{Lee11}
for evidence of Keplerian disks
around Class I object TMC1A and HH 111, respectively). Generally the
problem in
studying the kinematics on disk scales in embedded protostars is that
many of the traditional line tracers are optically thick in
the larger scale envelope. \cite{evolpaper}, for example, showed that
the emission from the (sub)millimeter transitions of HCO$^+$ would be
optically thick on scales of $\sim$100~AU for envelopes with masses
larger than 0.1~$M_\odot$. An alternative is to trace less
abundant isotopologues. Recently one embedded protostar, L1527, was
found to show Keplerian rotation in $^{13}$CO/C$^{18}$O 2--1
(\citealt{tobin12l1527}, see also \citealt{Murillo13} for VLA1623A); this result was subsequently strengthened
by ALMA observations (N. Ohashi, priv.\ comm., see also
  http://www.almasc.org/upload/presentations/DC-05.pdf).  Another
example of Keplerian motions is found in the protostellar binary
L1551-NE (a borderline Class 0/I source), in $^{13}$CO/C$^{18}$O 2--1
\citep{takakuwa12}. Large circumstellar disks are inferred around a number of high-mass young stellar objects, including G31.41+0.31 (\citealt{Cesaroni94}),
IRAS 20126+4104 (\citealt{Cesaroni97}), IRAS 18089-1732 (\citealt{Beuther04}), IRAS 16547-4247\linebreak (\citealt{Franco-Hernandez09}) and IRAS 18162-2048 (\citealt{Fernandez-Lopez11}). Whether they are rotationally supported remains uncertain.

\cite{brinch09} investigated the dynamics of the deeply embedded
protostar NGC~1333-IRAS2A in subarcsecond resolution observations of
HCN (4--3) and the same line of its isotopologue H$^{13}$CN\@. Through
detailed line radiative transfer modeling they showed that the
$\sim$~300~AU compact structure seen in dust continuum was in fact
dominated by infall rather than rotation.

\cite{pineda12} presented some of the first ALMA observations of the
deeply embedded protostellar binary IRAS~16293-2422. These
sensitive observations revealed a velocity gradient across one
component in the binary, IRAS16293A, in lines of the complex organic
molecule methyl formate. However, this velocity gradient does not
reflect Keplerian rotation and does not require a central mass beyond
the envelope mass enclosed on the same scale. This is also true
for the less dense gas traced by lines of the rare C$^{17}$O and
C$^{34}$S isotopologues in extended Submillimeter Array (eSMA) observations (0.5$''$ resolution;
\citealt{Favre14}).

The above observations paint a picture of complex structure of the
material on small scales around low-mass protostellar systems. They
raise a number of important potential implications, which we discuss
next.

\subsection{Implications and Outlook}
\label{ObsOutlook}

\subsubsection{Protostellar Mass Evolution}
\begin{figure}
  \resizebox{\hsize}{!}{\includegraphics{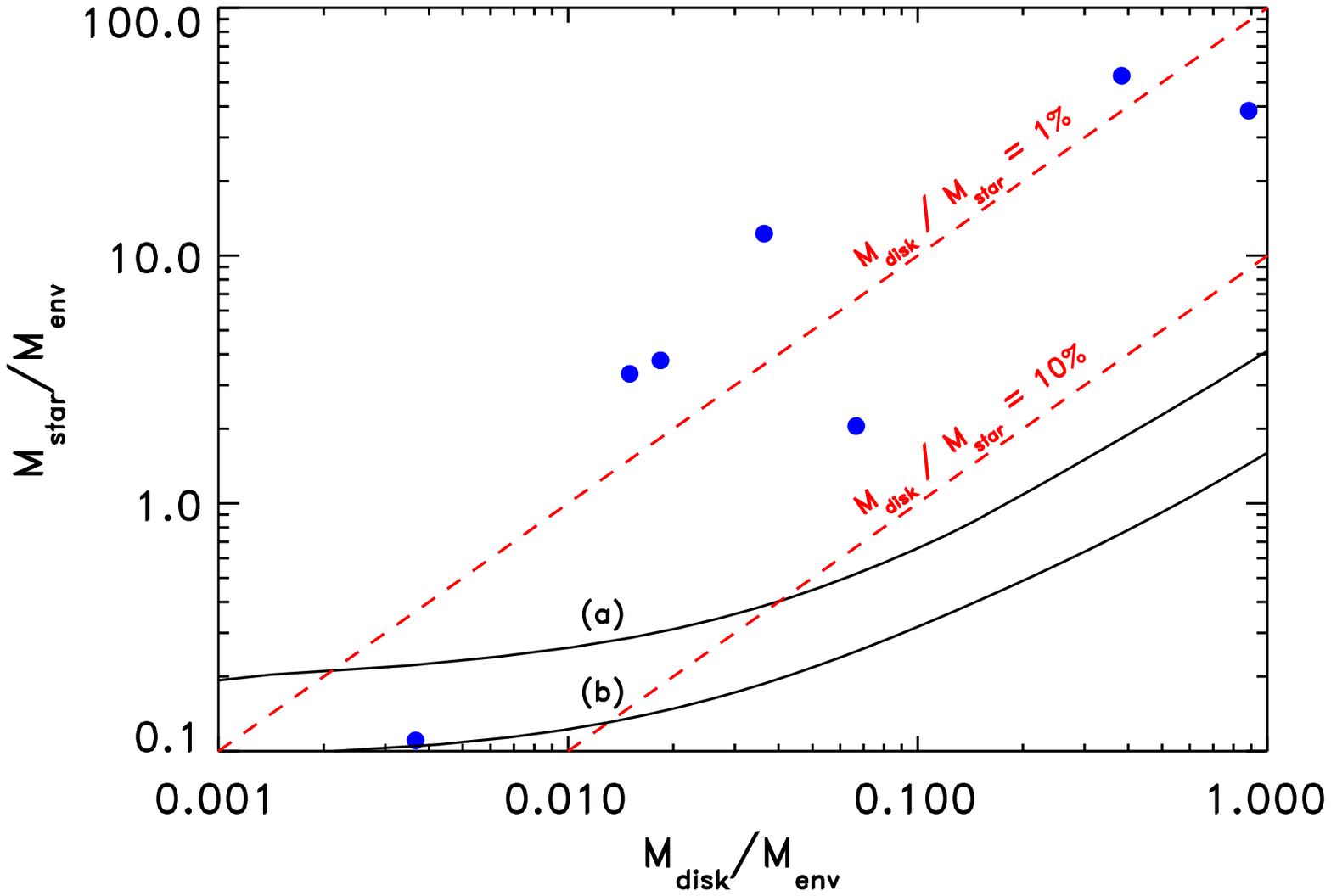}}
  \caption{Updated version of Fig.~18 from \cite{evolpaper}. Predicted
    stellar mass, $M_{\rm star}$, vs.\ disk mass, $M_{\rm disk}$, both
    measured relative to the envelope mass $M_{\rm env}$ in the models
    of \cite{visser09} with $\Omega_0 = 10^{-14}$~s$^{-1}$ and $c_s$
    of (a) 0.19~km~s$^{-1}$ and (b) 0.26~km~s$^{-1}$ (solid lines).
    The sources for which stellar, disk and envelope masses are
    measured are shown with blue dots
    \citep{evolpaper,takakuwa12,tobin12l1527}. Finally, the dashed
    lines indicate $M_{\rm disk}/M_{\rm star}$ ratios of 1\% (upper)
    and 10\% (lower).}\label{masscomparison}
\end{figure}

An important constraint on the evolution of protostars is how the bulk
mass is transported and accreted from the larger scales through the
circumstellar disks onto the central stars (\S~\ref{Disk2Planet}). An
important diagnostic
from the above observations is to compare the disk masses --- either
from dust continuum or line observations (taking into account the
caveats about the dust properties and/or chemistry) to stellar masses
inferred, e.g., from the disk dynamical profiles. \cite{evolpaper}
compared these quantities for a sample of predominantly Class I young
stellar objects with well-established disks: an updated version of this
figure including recent measurements of dynamical masses for
additional sources \citep{tobin12l1527,takakuwa12} is shown in
Fig.~\ref{masscomparison}. These measurements are compared to standard
semi-analytic models for collapsing rotating protostars
\citep{visser09}: these models typically underestimate the stellar
masses relative to the disk masses. These simple models of collapse of
largely spherical cores are likely inapplicable on larger scales where
filamentary structures are sometimes observed \citep{Tobin10,lee12}.
Still, this comparison
illustrates a potential avenue to explore when observations of a large sample of
embedded protostars become available in the ALMA era; it provides direct measures of
the accretion rates that more sophisticated numerical simulations
need to reproduce.

\subsubsection{Grain growth}
Multi-wavelength continuum observations in the millimeter and
submillimeter are also
interesting for studying the grain properties close to the newly
formed protostars. In more evolved circumstellar
disks a flattened slope of the spectral energy distribution at
millimeter or longer wavelengths is taken as evidence that significant
grain growth has taken place (see, e.g., \citealt{nattappv} for a
review).  On larger scales of protostellar envelope where
the continuum emission is optically thin, ISM-like dust
would result in submillimeter spectral slopes of
3.5--4. The more compact dust components observed at few hundred
AU scales have lower spectral indices of 2.5--3.0
\citep{prosacpaper,kwon09,chiang12}, either indicating that some
growth of dust grains to millimeter sizes has occurred or that the
compact components are optically thick.  At least in a few cases where
the dust emission is clearly resolved the inferred spectral slopes are
in agreement with those observed for more evolved T~Tauri stars,
indicating that dust rapidly grows to millimeter sizes
\citep{ricci10}.  Assessing
the occurrence of grain growth during the embedded stages is important
for not only understanding the formation of the seeds for
planetesimals (\S~\ref{Disk2Planet}), but also evaluating the
non-ideal MHD effects in magnetized core collapse and disk
formation, since they depend on the ionization level, which in turn
is strongly affected by the grain size distribution (\S~\ref{DiskTheory}).

\subsubsection{Chemistry}
The presence of disk-like structures on hundred AU scales may also
have important implications for the chemistry in those regions. The
presence of a disk may change the temperature, allowing molecules to
freeze-out again which would otherwise stay in the gas-phase.  Water and
its isotopologues are a particularly clear example of these effects:
for example in observations of the H$_2^{18}$O isotopologue toward the
centers of a small sample of Class~0 protostars, \cite{iras4b_h2o} and
\cite{persson12} found lower H$_2$O abundances than expected in the
typical gas-phase, consistent with a picture in which a significant
fraction of the material at scales $\lesssim$~100~AU has temperatures
lower than $\sim$~100~K\@.
This complicates the
interpretation of the chemistry throughout the envelope. For example, it
makes extrapolations of envelope physical and abundance structures
from larger scales invalid \citep{visser13}. An
ongoing challenge is to better constrain the physical structure of the
protostellar envelopes and disks on these scales. This must be done
before significant progress can be made on understanding the initial
conditions for chemical evolution in protoplanetary disks.
The challenge is even more formidable for massive stars, although
 \cite{Isokoski13} did not find any chemical differentiation between
massive stars with and without disk-like structures.

\subsubsection{Linking Observations and Theory}

Observations suggest that, in principle, there is typically more than enough angular
momentum on the core scale to form large, $10^2$~AU scale rotationally
supported disks. The common presence of fast jets
around deeply embedded protostars implies that the formation of
such disks has begun early in the process of star formation.
There is, however, currently little direct evidence that large, well-developed,
Keplerian disks are prevalent around Class 0 protostars, as one may
naively expect based on angular momentum conservation during
hydrodynamic core collapse. The paucity of large, early disks
indicates that disk formation is not as straightforward as generally
expected. The most likely reason is that star-forming cores are
observed to be significantly magnetized, and magnetic fields are known
to interact strongly with rotation. They greatly affect, perhaps even
control, disk formation, as we show next.

\section{Theory of Magnetized Disk Formation}
\label{DiskTheory}

How disks form has been a long-standing theoretical problem in star
formation. Early work on this topic was reviewed by
\citet{Bodenheimer95} and \citet{Boss98}. Both
reviews listed a number of unsolved problems. Topping both lists was
the effect of the magnetic field, which turns out to present a
formidable obstacle to disk formation.
Substantial progress has been made in recent years in overcoming this
obstacle, especially through Ohmic dissipation, field-rotation
misalignment, and turbulence. This progress will be summarized below.

\subsection{Magnetic Braking Catastrophe in Ideal MHD Limit}
\label{Catastrophe}

The basic difficulty with disk formation in magnetized dense cores can
be illustrated using analytic arguments in the strict ideal MHD limit
where the magnetic field lines are perfectly frozen into the core
material. In
this limit, as a finite amount of mass is accreted onto the central
object (the protostar), a finite amount of magnetic flux will be
dragged into the object as well. The magnetic flux accumulated at the
center forms a magnetic split monopole, with the field lines fanning
out radially (\citealt{Galli06}; see the sketch in their Fig.\ 1). As
a result, the magnetic field strength increases rapidly with
decreasing distance to the center, as $B\propto r^{-2}$. The magnetic
energy density, which is proportional to the field strength squared,
increases with decreasing radius even more rapidly, as $E_B\propto
r^{-4}$. This increase is faster than, for example, the energy density
of the accretion flow, which can be estimated approximately from
spherical free-fall collapse as $E_{\rm ff}\propto r^{-5/2}$. As the
infalling material approaches the central object, it will become
completely dominated by the magnetic field sooner or later. The
strong magnetic field at small radii is able to remove all of the
angular momentum in the collapsing flow, leading to the so-called
``magnetic braking catastrophe'' for disk formation
(\citealt{Galli06}). The braking occurs naturally in a magnetized
collapsing core, because the faster rotating matter that has already
collapsed closer to the rotation axis remains connected to the
more slowly rotating material at larger (cylindrical) distances
through field lines. The differential rotation generates a fieldline
twist that brakes the rotation of the inner, faster rotating part
and transports its angular momentum outward (see
\citealt{Mouschovias79,Mouschovias80}
for analytic illustrations of magnetic braking).

The catastrophic braking of disks in magnetized dense cores in the
ideal MHD limit was also found in many numerical as well as
semi-analytic calculations. \linebreak \citet{Krasnopolsky02} were the
first to show semi-analytically, using the so-called ``thin-disk''
approximation,
that the formation of rotationally supported disks (RSDs hereafter)
can be
suppressed if the efficiency of magnetic braking is large enough.
However, the braking efficiency
was parametrized rather than computed self-consistently.
Similarly, \citet{Dapp10} and \citet{Dapp12} demonstrated
that, in the absence of any magnetic diffusivity, a magnetic
split-monopole is produced at the center and RSD formation is
suppressed, again under the thin-disk approximation.

Indirect evidence for potential difficulty with disk formation in
ideal MHD simulations came from \citet{Tomisaka00}, who
studied the (2D) collapse of a rotating, magnetized dense core
using a nested grid under the assumption of axisymmetry. He found
that, while there is little magnetic
braking during the phase of runaway core collapse leading up to
the formation of a central object, once an outflow is launched,
the specific angular momentum of the material at the highest
densities is reduced by a large factor (up to $\sim 10^4$) from
the initial value. The severity of the magnetic braking and its
deleterious effect on disk formation were not fully appreciated
until \citet{Allen03b} explicitly demonstrated that the
formation of a large, numerically resolvable, rotationally
supported disk was completely suppressed in 2D by a moderately
strong magnetic field (corresponding to a dimensionless
mass-to-flux ratio $\lambda$ of several) in an initially self-similar,
rotating, magnetized, singular isothermal toroid \citep{LiShu96}.
They identified two key ingredients behind the efficient braking
during the accretion phase: (1) concentration of the field lines at small
radii by the collapsing flow, which increases the field strength,
and (2) the fanning out of field lines due to equatorial pinching,
which increases the lever arm for magnetic braking.

The catastrophic braking that prevents the formation of RSDs during
the protostellar accretion phase has been confirmed in several
subsequent 2D and 3D ideal MHD simulations
\citep{Mellon08,Hennebelle08,Duffin09,Seifried12,SantosLima12}.
\citet{Mellon08}, in particular, formulated the disk formation
problem in the same way as \citet{Allen03b}, by adopting a
self-similar rotating, magnetized, singular isothermal toroid as
the initial configuration. Although idealized, the adopted initial
configuration has the advantage that the subsequent core collapse
should remain self-similar. The self-similarity provides a useful
check on the correctness of the numerically obtained solution.
They found that the disk formation was suppressed by a field as weak
as $\lambda=13.3$.

\citet{Hennebelle08} carried out 3D simulations of the
collapse of a rotating dense core of uniform density and magnetic
field into the protostellar accretion phase using an ideal MHD AMR
code (RAMSES). They found that the formation of a RSD is suppressed as
long as $\lambda$ is of order 5 or less. However, the $\lambda=5$
case in \citeauthor{Price07}'s (\citeyear{Price07}) SPMHD simulations
appears to have formed a small disk (judging from the column density
distribution). It is unclear whether the disk is rotationally
supported or not, since the disk rotation rate was not given in
the paper. Furthermore, contrary to the grid-based simulations, there
appears little, if any, outflow driven by twisted field lines in
the SPH simulations, indicating that the efficiency of magnetic
braking is underestimated (prominent outflows are produced,
however, in other SPMHD simulations, e.g., \citealt{Burzle11} and \citealt{Price12} ). Another apparently discrepant result
is that of \citet{Machida11}. They managed to form a large,
$10^2$-AU scale, rotationally supported disk for a very strongly
magnetized core of $\lambda=1$ even in the ideal MHD limit
(their Model 4) using a nested grid and sink particle. This
contradicts the results from other
simulations and semi-analytic calculations.

To summarize, both numerical simulations and analytic arguments
support the notion that, in the ideal MHD limit, catastrophic
braking makes it difficult to form rotationally supported disks
in (laminar) dense cores magnetized to a realistic level (with
a typical $\lambda$ of a few). In what follows, we will explore
the potential resolutions that have been proposed in the
literature to date.

\subsection{Non-ideal MHD Effects}

Dense cores of molecular clouds are lightly ionized (with a typical
electron fractional abundance of order $10^{-7}$; \citealt{Bergin07}).
As such, the
magnetic field is not expected to be perfectly frozen into the bulk
neutral material. There are three
well known non-ideal MHD effects that can in principle break the
flux-freezing condition that lies at the heart of the magnetic
braking catastrophe in the strict ideal MHD limit. They are ambipolar
diffusion, the Hall effect, and Ohmic dissipation (see \citealt{Armitage11}
for a review). Roughly speaking,
in the simplest case of an electron-ion-neutral medium, both ions and
electrons are well tied to the magnetic field in the ambipolar
diffusion regime. In the Hall regime, electrons remain well tied to
the field, but not ions. At the highest densities, both electrons and
ions are knocked off the field lines by collisions before they finish
a complete gyration; in such a case, Ohmic dissipation dominates.
This simple picture is complicated by dust grains, whose size
distribution in dense cores is relatively uncertain (see
\S~\ref{ObsOutlook}), but which can become the dominant charge
carriers. Under typical
cloud conditions, ambipolar diffusion dominates over the other two effects
at densities typical of cores (e.g., \citealt{Nakano02}, \citealt{Kunz10}). It is the most
widely studied non-ideal MHD effect in the context of core formation
and evolution in the so-called ``standard'' picture of low-mass star
formation out of magnetically supported clouds
\citep{Nakano84,Shu87,Mouschovias99}. It is the effect that we will
first concentrate on.

\begin{figure}[ht]
\resizebox{\hsize}{!}{\includegraphics{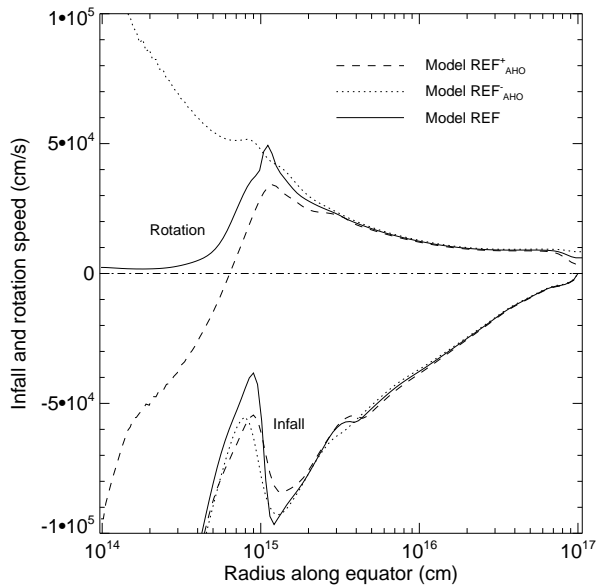}}
\caption{Infall and rotation speeds along the equatorial plane of a
  collapsing rotating, magnetized dense core during the protostellar
  mass accretion phase for three representative models of \citet{Li11}.
  Model REF (solid lines) includes only ambipolar
  diffusion, whereas the other two include all three non-ideal MHD
  effects, especially the Hall effect. The initial magnetic field
  and rotation axis are in the
  same direction in one model (REF$^+$, dashed lines) and in opposite
  directions in the other (REF$^-$, dotted).}\label{nonideal}
\end{figure}

Ambipolar diffusion enables the magnetic field lines that are tied
to the ions to drift relative to the bulk neutral material. In the
context of disk formation, its most important effect is to
redistribute the magnetic flux that would have been dragged into
the central object in the ideal MHD limit to a circumstellar region
where the magnetic field strength is greatly enhanced \citep{LiMcKee96}.
Indeed, the enhanced circumstellar magnetic field is strong
enough to drive a hydromagnetic shock into the protostellar accretion flow
\citep{LiMcKee96,Ciolek98,Contopoulos98,Krasnopolsky02,Tassis07,Dapp12}.
\citet{Krasnopolsky02} showed semi-analytically, using the 1D thin-disk
approximation,
that disk formation may be suppressed in the strongly magnetized
post-shock region if the magnetic braking is efficient enough. The
braking efficiency, parametrized in \citet{Krasnopolsky02},
was computed self-consistently in the 2D (axisymmetric) simulations
of \citet{Mellon09}, which were performed under the usual
assumption of ion density proportional to the square root of
neutral density.  3D simulations of AD were performed by \citet{Duffin09}
using a specially developed, single fluid AMR code \citep{Duffin08} as
well as by a two fluid SPH code \citep{Hosking04}.
\citet{Mellon09} found that ambipolar
diffusion does not weaken the magnetic braking enough to allow
rotationally supported disks to form for realistic levels of
cloud core magnetization and cosmic ray ionization rate. In many
cases, the magnetic braking is even enhanced. These findings were
strengthened by \citet{Li11}, who computed the ion density
self-consistently using the simplified chemical network of Nakano
and collaborators \citep{Nakano02,Nishi91} that
includes dust grains. An example of their simulations is shown
in Fig.\ \ref{nonideal}. It shows clearly the rapid slowdown
of the infalling material near the ambipolar diffusion-induced
shock (located at a radius $\sim 10^{15}$~cm in this particular
example) and the nearly complete braking of the rotation in the
post-shock region, which prevents the formation of a RSD\@. The
suppression of RSD is also evident from the fast, supersonic
infall close to the central object. We should note that RSD formation
may still be possible if the cosmic ray ionization rate can be
reduced well below the canonical value of $10^{-17}$~$s^{-1}$
\citep{Mellon09}, through for example the magnetic mirroring
effect, which may turn a large fraction of the incoming cosmic
rays back before they reach the disk-forming region \citep{Padovani13}.

As the density increases, the Hall effect tends to become more
important (the exact density for this to happen depends on the
grain size distribution).
It is less explored than ambipolar diffusion in the star formation
literature. A unique feature of this effect is that it can
actively increase the angular momentum of a collapsing, magnetized
flow through the so-called
``Hall spin-up'' \citep{Wardle99}. In the simplest case of
electron-ion-neutral fluid, the spin-up is caused by the current
carriers (the electrons) moving in the azimuthal direction, generating
a magnetic torque through field twisting; the toroidal current is
produced by gravitational collapse, which drags the poloidal field
into a pinched, hourglass-like configuration. The Hall spin-up was
studied numerically by \citet{Krasnopolsky11}
and semi-analytically by \citet{Braiding12a,Braiding12b}.
\citet{Krasnopolsky11} showed that a rotationally supported disk can
form even in an initially {\it non-rotating} core, provided that
the Hall coefficient is large enough. Interestingly, when the
direction of the initial magnetic field in the core is flipped,
the disk rotation is reversed. This reversal of rotation is also
evident in Fig.\ \ref{nonideal}, where the Hall effect spins up
the nearly non-rotating material in the post-AD shock region
to highly supersonic speeds, but in different directions
depending on the field orientation.
The Hall effect, although dynamically significant, does not
appear capable of forming a rotationally supported disk under
typical dense core conditions according to \citet{Li11}. This
inability is illustrated in Fig.\ \ref{nonideal}, where the equatorial material
collapses supersonically on the $10^2$-AU scale even when the Hall
effect is present.

Ohmic dissipation becomes the dominant nonideal MHD effects at high
densities (e.g., \citealt{Nakano02}). It has been investigated by
different groups in connection with disk formation. \citet{Shu06}
studied semi-analytically the effects of a spatially uniform
resistivity on the magnetic field structure during the protostellar
mass accretion phase. They found that, close to the central object,
the magnetic field decouples from the collapsing material and
becomes more or less uniform. They suggested that a rotationally
supported disk may form in the decoupled region, especially if the
resistivity is higher than the classic (microscopic) value. This
suggestion was confirmed by
\citeauthor{Krasnopolsky10} (\citeyear{Krasnopolsky10}; see also
\citealt{SantosLima12}), who found numerically that a large,
$10^2$AU-scale, Keplerian disk can form around a $0.5$~M$_\odot$ star,
provided that the resistivity is of order $10^{19}$~cm$^2$~s$^{-1}$
or more; such a resistivity is significantly higher than the classic
(microscopic) value over most of the density range relevant to disk
formation.

\citet{MachidaMatsumoto11} and \citet{Machida11} studied disk
formation in magnetized cores including only the classic value of
resistivity estimated from \citeauthor{Nakano02}'s (\citeyear{Nakano02})
numerical results.
The former study found that
a relatively small, 10~AU-scale, rotationally supported disk formed
within a few years after the formation of the stellar core. Inside
the disk, the density is high enough for magnetic decoupling to occur
due to Ohmic dissipation. This work was extended to much later times
by \citet{Machida11}, who included a central sink region in the
simulations. They concluded that the small RSD can grow to large,
$10^2$-AU size at later times, especially after the most of the
envelope material has fallen onto the disk and the central object. A
caveat, pointed out by
\citeauthor{Tomida13} (\citeyear{Tomida13}; see also \citealt{Dapp10}),
is that they used a form of induction equation that is,
strictly speaking, inappropriate for the non-constant resistivity
adopted in their models; it may generate magnetic monopoles that are
subsequently cleaned away using
the Dedner's method \citep{Dedner02}. This deficiency was corrected
in \citet{Tomida13}, who carried out radiative MHD simulations of
magnetized core collapse to a time shortly ($\sim 1$~year) after the
formation of the second (protostellar) core. They found that the
formation of a (small, AU-scale) rotationally supported disk was
suppressed by magnetic braking in the ideal MHD limit but was enabled
by Ohmic dissipation at this early time; the latter result is in
qualitative agreement with \citet{MachidaMatsumoto11} and
\citet{Machida11}, although it remains to be seen how the small disks in
\citet{Tomida13}'s simulations evolve further in time.

\citet{Dapp10} studied the effects of Ohmic dissipation on disk
formation semi-analytically, using the ``thin-disk'' approximation
for the mass distribution and an approximate treatment of magnetic
braking. The approximations enabled them to follow the formation of
both the first and second core. They found that a small, sub-AU,
rotationally supported disk was able to form soon after the formation
of the second core in the presence of Ohmic dissipation; it was
suppressed in the ideal MHD limit, in agreement with the later 3D
simulations of \citet{Tomida13}. This work was extended by \citet{Dapp12}
to include a set of self-consistently computed charge
densities from a simplified chemical network and ambipolar diffusion.
They showed that their earlier conclusion that a small, sub-AU scale,
RSD is formed through Ohmic dissipation holds even in the
presence of a realistic level of ambipolar diffusion. This conclusion
appears reasonably secure in view of the broad agreement between the
semi-analytic work and numerical simulations. When and how such disks
grow to the much larger, $10^2$AU-scale, size deserve to be explored
more fully.

\subsection{Magnetic Interchange Instabilities}

The formation of a large-scale RSD in a magnetized core is made
difficult by the accumulation of magnetic flux near the
accreting protostar. As discussed earlier, this is especially true
in the presence of a realistic level of ambipolar diffusion, which
redistributes the magnetic flux that would have been dragged into
the central object to the circumstellar region (Ohmic dissipation
has a similar effect, see \citealt{Li11} and \citealt{Dapp12}).
The result of the flux redistribution is the creation of a
strongly magnetized region close to the protostar where the infall
speed of the accreting flow is slowed down to well below the free-fall
value (i.e., it is effectively held up by magnetic tension against
the gravity of the central object), at
least in 2D (assuming axisymmetry). It has long been suspected
that such a magnetically supported structure would become unstable
to interchange instabilities in 3D \citep{LiMcKee96,Krasnopolsky02}.
Recent 3D simulations have shown that this is indeed the case.

Magnetic interchange instability in a protostellar accretion flow
driven by flux redistribution was first studied in detail by \citet{Zhao11}.
They treated the flux redistribution through a sink
particle treatment: when the mass in a cell is accreted onto a sink
particle, the magnetic field is left behind in the cell (see also
\citealt{Seifried11} and \citealt{Cunningham12}); it is a
crude representation of the matter-field decoupling expected at
high densities (of order $10^{12}$~cm$^{-3}$ or higher; \citealt{Nakano02};
\citealt{Kunz10}). The decoupled flux piles up
near the sink particle, leading to a high magnetic pressure that
is released through the escape of field lines along the directions
of least resistance. As a result, the magnetic flux dragged into
the decoupling region near the protostar along some azimuthal
directions is advected back out along other directions in highly
magnetized, low-density, expanding regions. Such regions are
termed DEMS (decoupling-enabled magnetic structure) by \citet{Zhao11};
they appear to be present in the formally ideal MHD
simulations of \citet{Seifried11}, \citet{Cunningham12}
and \citet{Joos12} as well.

\begin{figure}[ht]
\resizebox{\hsize}{!}{\includegraphics{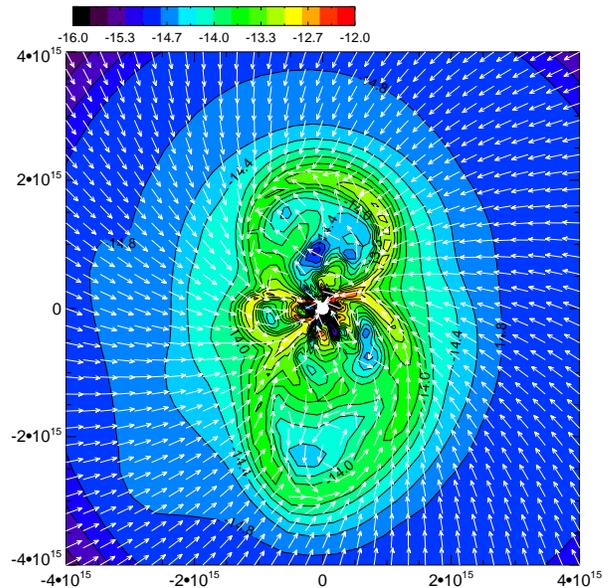}}
\caption{An example of the inner protostellar accretion flow driven
unstable by magnetic flux redistribution (taken from
\citealt{Krasnopolsky12}).
Plotted are the distribution of logarithm of density
(in units of g~cm$^{-3}$) and velocity
field on the equatorial plane (the length is in units of cm). The
expanding, low-density, regions near the center
are the so-called ``decoupling-enabled magnetic structure'' (DEMS)
that are strongly magnetized. They present a formidable barrier to disk
formation.}\label{DEMS}
\end{figure}

\citet{Krasnopolsky12} improved upon the work of \citet{Zhao11}
by including two of the physical processes that can lead
to magnetic decoupling: ambipolar diffusion and Ohmic dissipation.
They found that the basic conclusion of \citet{Zhao11} that the inner
part of the
protostellar accretion flow is driven unstable by magnetic flux
redistribution continues to hold in the presence of realistic
levels of non-ideal MHD effects (see Fig.\ \ref{DEMS} for an
illustrative example). The magnetic flux accumulated
near the center is transported outward not only diffusively by
the microscopic non-ideal effects, but also advectively through
the bulk motions of the strongly magnetized expanding regions
(the DEMS) generated by the instability. The advective flux
redistribution in 3D lowers the field strength at small radii
compared to the 2D (axisymmetric) case where the instability is
suppressed. It makes the magnetic braking less efficient and the
formation of a RSD easier in principle. In practice, the magnetic
interchange instability does not appear to enable the formation
of rotationally supported disks by itself, because the highly
magnetized DEMS that it creates remain trapped at relatively
small distances from the protostar by the protostellar accretion
flow (see Fig.\ \ref{DEMS}); the strong magnetic field inside
the DEMS blocks the accretion flow from rotating freely around
the center object to form a complete disk.

\subsection{Magnetic Field-Rotation Misalignment}

\begin{figure}
\resizebox{\hsize}{!}{\includegraphics{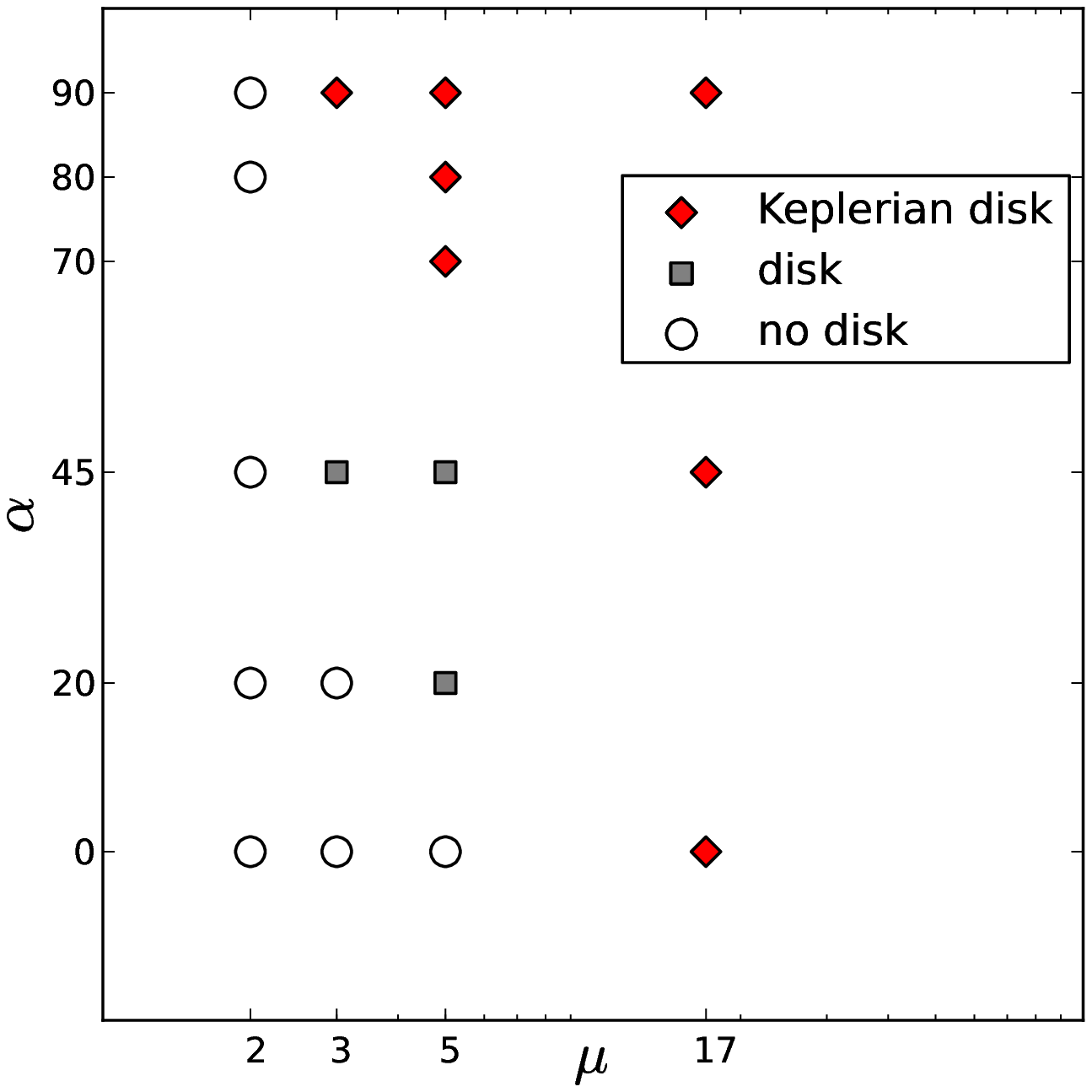}}
\caption{Parameter space for disk formation according to \citet{Joos12}.
  The parameter $\alpha$ is the angle between the magnetic
  field and rotation axis, and $\mu$ is the dimensionless mass-to-flux
  ratio of the dense core (denoted by $\lambda$ in the rest of the
  article). The diamond denotes disks with Keplerian rotation profile,
  the square those with flat rotation curve, and circle the cases with
  no significant disk.}\label{Joos}
\end{figure}

Misalignment between the magnetic field and rotation axis as a way to
form large RSDs has been explored extensively by Hennebelle and
collaborators (\citealt{Hennebelle09,Ciardi10,Joos12};
see also \citealt{Machida06,Price07}, and \citealt{Boss13}).
The misalignment is expected if the angular
momenta of dense cores are generated through turbulent motions
\citep[e.g.,][]{Burkert00,Seifried12b,Myers13,Joos13}.
Plausible observational
evidence for it was recently uncovered by \citet{Hull13} using
CARMA, who
found that the distribution of the angle between the magnetic field
on the $10^3$AU-scale and the bipolar outflow axis (taken as a proxy
for the rotation axis) is consistent with being random. If true, it
would imply that in half of the sources the two axes are misaligned
by an angle greater than $60^\circ$. \citet{Joos12} found that
such a large misalignment enables the formation of RSDs in moderately
magnetized dense cores with a dimensionless mass-to-flux ratio
$\lambda$ of $\sim 3$--$5$; RSD formation is suppressed in such cores
if the magnetic field and rotation axis are less misaligned
(see Fig.\ \ref{Joos}). They
attributed the disk formation to a reduction in the magnetic braking
efficiency induced by large misalignment. In more strongly magnetized
cores with $\lambda \lesssim 2$, RSD formation is suppressed
independent of the misalignment angle, whereas in very weakly
magnetized cores RSDs are formed for all misalignment angles.

Based on the work of \citet{Hull13} and \citet{Joos12},
\citet{Krumholz13} estimated that the field-rotation misalignment
may enable the formation of large RSDs in $\sim 10$--$50\%$ of dense
cores. If the upper range is correct, the misalignment would go a
long way toward solving the problem of excessive magnetic braking in
protostellar disk formation.

\citet{Li13} carried out simulations similar to those of \citet{Joos12},
except for the initial conditions. They confirmed the
qualitative result of \citet{Joos12} that the field-rotation
misalignment is conducive to disk formation. In particular, large misalignment
weakens the strong outflow in the aligned case and is a key reason
behind the formation of RSDs in relatively weakly magnetized cores.
For more strongly magnetized cores with $\lambda \lesssim 4$, RSD
formation is suppressed independent of the degree of misalignment.
This threshold value for the mass-to-flux ratio is about a factor
of 2 higher than that obtained by \citet{Joos12}. The difference
may come, at least in part, from the different initial conditions
adopted: uniform density with a uniform magnetic field for \citet{Li13}
and a centrally condensed density profile with a
nonuniform but unidirectional field for \citet{Joos12}; the
magnetic braking is expected to be more efficient at a given (high)
central density for the former initial configuration, because
its field lines would become more pinched, with a longer lever arm
for braking. Whether there are other factors that contribute
significantly to the above discrepancy remains to be determined.

If the result of \citet{Li13} is correct, then a dense core
must have both a large field-rotation misalignment {\it and} a
rather weak magnetic field in order to form a RSD\@. This dual
requirement would make it difficult for the misalignment alone
to enable disk formation in the majority of dense cores, which
are typically rather strongly magnetized according to \citeauthor{Troland08}
(\citeyear{Troland08}, with a median mass-to-flux ratio of $\lambda\sim
2$). In a more recent study, \citet{Crutcher10} argued, based on
  Bayesian analysis, that a fraction of dense cores could be very weakly
  magnetized, with a dimensionless mass-to-flux ratio $\lambda$ well above
  unity (see \citealt{Bertram12} for additional arguments for
  weak field, including field reversal). However, since the median
  mass-to-flux ratio remains unchanged
  for the different distributions of the total field strength assumed
  in \citeauthor{Crutcher12}'s (\citeyear{Crutcher12})
  Bayesian analysis, it is unlikely for
  the majority of dense cores to have $\lambda$ much greater than the
  median value of $2$. For example, \citet{Li13} estimated the
  fraction of dense cores with $\lambda > 4$ at $\sim 25\%$.
There is also concern that the random distribution of the
field-rotation misalignment angle found by \citet{Hull13}
on the $10^3$~AU scale may not be representative of the
distribution on the larger core scale. Indeed, \citet{Chapman13}
found that the field orientation on the core scale (measured
using a single dish telescope) is within $\sim 30^\circ$ of the
outflow axis for 3 of the 4 sources in their sample (see also
\citealt{Davidson11}); the larger angle measured in the remaining
source may be due to projection effects because its outflow axis
lies close to the line of sight. If the result of \citet{Chapman13}
is robust and if the outflow axis reflects the rotation
axis, dense cores with large misalignment between the magnetic
and rotation axes would be rare. In such a case, it would be even
less likely for the misalignment to be the dominant mechanism for
disk formation.

\subsection{Turbulence}
\label{turbulence}

\begin{figure}
\resizebox{\hsize}{!}{\includegraphics[trim=243 347 486 0,clip]{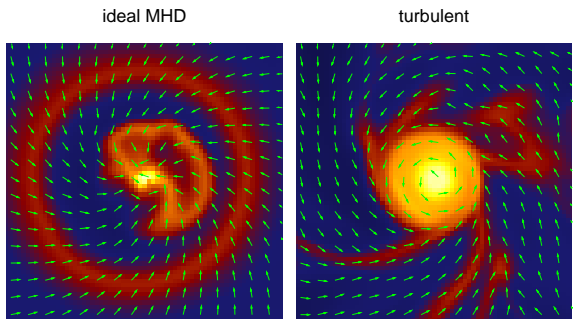}\includegraphics[trim=729 347 0 0,clip]{SantosLima.eps}}
\caption{Accretion of rotating, magnetized material onto a
  pre-existing central object with (right panel) and without (left)
  turbulence (adapted from \citealt{SantosLima12}).
  The formation of a nearly Keplerian disk is clearly suppressed in the
  laminar case (left panel), by excessive magnetic braking, but is
  enabled by turbulence (right).}\label{SantosLima}
\end{figure}

Turbulence is a major ingredient for star formation (see reviews
by, e.g., \citealt{MacLow04} and \linebreak \citealt{McKee07}). It
can generate local angular
momentum by shear flows and form highly asymmetric dense cores
(see results from Herschel observations, e.g., \citealt{Menshchikov10}
and \citealt{Molinari10}).
There is increasing evidence that it also promotes
RSD formation. \citet{SantosLima12} contrasted
the accretion of turbulent and laminar magnetized gas onto a
pre-existing central star, and found that a nearly Keplerian disk
was formed in the turbulent but not laminar case (see
Fig.\ \ref{SantosLima}).  The simulations were carried out at a
relatively low resolution (with a rather large cell size of 15.6~AU;
this was halved, however, in \citealt{SantosLima13}, who found
similar results), and turbulence was driven to an rms Mach number of
$\sim 4$, which may
be too large for low-mass cores. Nevertheless, the beneficial effect
of turbulence on disk formation is clearly demonstrated. They attributed
the disk formation to the turbulence-induced outward diffusion of
magnetic flux, which reduces the strength of the magnetic field
in the inner, disk-forming, part of the accretion flow.
Similar results of disk formation in turbulent cloud cores are
presented by \citet{Seifried12b} and \citet{Seifried13}, although these
authors attribute their findings to different mechanisms.
They argued that the turbulence-induced
magnetic flux loss is limited well outside their disks, based on
the near constancy of an approximate mass-to-flux ratio computed
on a sphere several times the disk size (with a $500$~AU radius).
They proposed instead that
the turbulence-induced tangling of field lines and strong local
shear are mainly responsible for the disk formation: the disordered
magnetic field weakens the braking and the shear enhances
rotation. Similarly, \citet{Myers13} also observed formation
of a nearly Keplerian disk in their radiative MHD simulation of a
turbulent massive (300~$M_\odot$) core, although they refrained
from discussing the origin of the disk in detail since it was not
the focus of their investigation.

\citet{Seifried13} extended their previous work to include
both low-mass and high-mass cores and both subsonic and supersonic
turbulence. They found disk formation in all cases. Particularly
intriguing is the formation of rotationally dominated disks in
the low-mass, subsonically turbulent cores. They argued that,
as in the case of massive core with supersonic turbulence of
\citet{Seifried12b}, such disks are not the consequence of
turbulence-induced magnetic flux loss, although such loss appears
quite severe on the disk scale, which may have contributed to
the long-term survival of the formed disk. While disks appear
to form only at sufficiently high mass-to-flux ratios $\lambda
\ge 10$ in ordered magnetic fields, disks in the turbulent
MHD simulations form at much lower and more realistic values
of $\lambda$. \citet{Seifried12b} found that $\lambda$ increases
gradually in the vicinity of the forming disk, which may have
more to do with the growing accreting mass relative to the
magnetic flux than with dissipative effects of the magnetic
field by turbulent diffusion or reconnection.

\citet{Joos13} investigated the effects of turbulence of various
strengths on disk formation in a core of intermediate mass ($5 M_\odot$).
They found that an initially imposed turbulence has two major effects.
It produces an effective diffusivity that enables magnetic flux to
diffuse outward, broadly consistent with the picture envisioned
in \citet{SantosLima12,SantosLima13}. It also generates a substantial
misalignment between the rotation axis and magnetic field direction
(an effect also seen in \citealt{Seifried12b} and \citealt{Myers13}).
Both of these
effects tend to weaken magnetic braking and make disk formation easier.
If the turbulence-induced magnetic diffusion is responsible, at least
in part, for the disk formation, then numerical effects would be
a concern. In the ideal MHD limit, the diffusion presumably comes
from turbulence-enhanced reconnections due to finite grid resolution.
Indeed, \citet{Joos13}
reported that their simulations did not appear to be fully converged,
with disk masses differing by a factor up to $\sim 2$ in higher
resolution simulations. The situation is further complicated by
numerical algorithms for treating magnetic field evolution, especially
those relying on divergence cleaning, which could introduce additional
artificial magnetic diffusion. To make further progress, it would be
useful to determine when and how the reconnections occur and exactly
how they lead to the magnetic diffusion that are apparent in the
simulations of \citet{Joos13}, \citet{SantosLima12}, Li et
al.\ (in preparation) and perhaps \citet{Seifried12b,Seifried13}.

\subsection{Other Mechanisms}

The magnetic braking catastrophe in disk formation would disappear
if the majority of dense cores are non-magnetic or only
weakly magnetized \citep[$\lambda \gsim 5$ or even $\lambda \gsim 10$, see e.g.,][]{Hennebelle09,
  Ciardi10, Seifried11}. However, such weakly magnetized cloud cores
are rather unlikely. Although, the recent study by \citet{Crutcher10}
indicates that some cloud cores might be highly supercritical, they
are certainly not the majority. Furthermore,
consider, for example, a typical core of 1~M$_\odot$ in
mass and $10^4$~cm$^{-3}$ in $H_2$ number density. To have a
dimensionless mass-to-flux ratio $\lambda \gtrsim 5$, its field
strength must be $B\lesssim 4.4$~$\mu$G, less than the median field
strength inferred for the atomic CNM ($\sim 6$~$\mu$G, \citealt{Heiles05}),
which is unlikely. We therefore expect the majority
of dense cores to have magnetic fields corresponding to $\lambda
\lesssim 5$ (in agreement with \citealt{Troland08}), which
are strong enough to make RSD formation difficult.

Another proposed solution is the depletion of the protostellar
envelope. The slowly rotating envelope acts as a brake on the
more rapidly rotating material closer to the central object that
is magnetically connected to it. Its depletion should promote RSD
formation. Indeed, \citet{Machida10} found that envelope depletion
is conducive to the formation of large RSDs toward the end of the
main accretion phase. This is in
line with the expectation of \citet{Mellon08}, who envisioned
that most of the envelope depletion is achieved through wind
stripping rather than accretion, as would be the case if the star
formation efficiency of individual cores is relatively low (say,
$\sim 1/3$, \citealt{Alves07}). Given the ubiquity of fast outflows,
their effects on envelope depletion and disk formation should
be investigated in more detail.

\subsection{Summary and Outlook}

The formation of rotationally supported disks turns out to be much
more complicated than envisioned just a decade ago. This is because
star-forming dense cores are observed to be rather strongly
magnetized in general (although the magnetization in a fraction
of them can be rather weak, see \citealt{Crutcher10} and
discussion above), with a magnetic energy typically much higher than the
rotational energy. The field strength is further amplified by core
collapse, which tends to concentrate the field lines in the region
of disk formation close to the protostar. Both analytic calculations
and numerical simulations have shown that the collapse-enhanced
(ordered) magnetic field can prevent the RSD formation through catastrophic
magnetic braking in the simplest case of ideal MHD limit, aligned
magnetic field and rotation axis, and no turbulence.  Ambipolar
diffusion, the Hall effect, and magnetic interchange instabilities
have profound effects on the dynamics of the inner protostellar
accretion flow, but they do not appear capable of forming RSDs
by themselves under typical conditions. Ohmic dissipation, on the
other hand, can enable the formation of at least small, AU-scale
disks at early times. How such disks evolve in the presence of
the instabilities and other non-ideal MHD effects remains to be
quantified. Magnetic field-rotation misalignment is conducive to
disk formation, but it is unlikely to enable the formation of
RSDs in the majority of dense cores, because of the dual requirement
of both a relatively weak magnetic field and a relatively large
tilt angle that may be uncommon on the core scale. Turbulence
appears to facilitate RSD formation in a number of numerical
simulations. It is possible that the turbulence-enhanced numerical
reconnection plays a role in the appearance of RSDs in these formally
ideal MHD simulations, although turbulence by itself could reduce
braking efficiency. The possible role of reconnection needs to be
better understood and quantified.

\section{Early Outflows}
\label{outflow}

\subsection{Introduction}

Generally, low mass young stellar objects are accompanied by highly
collimated optical jets \citep[see e.g.,][]{Cabrit97, Reipurth01}
whereas high mass stars are often obscured, hard to observe, and
until recently thought to drive much less collimated outflows
\citep[e.g.,][and discussion below]{Shepherd96a,
  Shepherd96b, Beuther05}. Nevertheless, there is strong evidence that
the underlying launching process is based on the same physical
mechanism, namely the magneto-rotational coupling: magnetic fields
anchored to an underlying rotor (e.g., an accretion disk) will carry
along gas which will be flung outwards \citep[the same mechanism could
also apply to galactic jets, e.g.,][]{Pudritz07, Pudritz09}. For
  example, \cite{Guzman12} concluded that collimated thermal radio
  jets are associated with high-mass young stellar objects, although
  for a relatively short time ($\sim 4\times 10^4$~yr).

With the seminal theoretical work by \citet{Blandford82} and \citet{Pudritz83}
the idea of magneto-centrifugally driven jets was
first established. It was shown that magnetic fields anchored
to the disk around a central object can lift off gas from the disk
surface. A magneto-centrifugally driven jet will be launched if the
poloidal component of magnetic field is inclined with respect to the
rotation axis by more than 30$^\circ$. Numerical
simulations of Keplerian accretion disks threaded by such a magnetic
field have shown that these jets are self-collimated and
accelerated to high velocities \citep[e.g.,][]{Fendt96, Ouyed97b}.
This driving mechanism, where the launching of the jet is connected to
the underlying accretion disk, predicts that jets rotate and carry
angular momentum off the disk. The first plausible observational
confirmation came from Hubble Space Telescope (HST) detection of rotational
signatures in the optical jet of DG Tauri (\citealt{Bacciotti02}; see,
however, \citealt{Soker05}, \citealt{Cerqueira06} and \citealt{Fendt11}
for different
interpretations of the observation). Further evidence is provided by
UV \citep{Coffey07} and IR observations \citep{Chrysostomou08}.

The most viable mechanism to launch jets and wider angle outflows from
accretion disks around YSOs is the coupling through magnetic fields
where the gas from the disk surface is accelerated by the Lorentz
force. Generally, this force can be divided into a magnetic tension
and a magnetic pressure term. In an axisymmetric setup, the magnetic
tension term is responsible for the magneto-centrifugal acceleration
and jet collimation via hoop stress. The magnetic pressure can also
accelerate gas off the underlying disk. These magnetic pressure driven
winds are sometimes known as {\it magnetic twist} \citep{Shibata85},
{\it plasma gun} \citep{Contopoulos95}, or {\it magnetic tower}
\citep{LyndenBell03}.

Protostellar disks around young stellar objects themselves are the
result of gravitational collapse of molecular cloud cores (see
Sec.\ \ref{DiskTheory}). Since the molecular clouds are permeated
by magnetic fields of varying strength and morphology
\citep[see e.g.,][]{Crutcher99, Beck01,Alves08}, there should be a
profound link between the collapse and magneto-rotationally driven
outflows.

There are still many unresolved problems concerning jets
and outflows. These include the details of the jet launching,
the driving of molecular outflows, the efficiency
of outflows around massive protostars, the influence of outflow
feedback on star formation, and how efficient
they are in clearing off the envelope material around the young
stellar object. Shedding light on the last
problem will also help determine whether there is a clear physical
link between the core mass function (CMF) and IMF (see the chapter by
Offner et al.\ in this volume).

In this section we summarize our knowledge of early jets and outflows
and discuss open questions.

\subsection{Jet Launching and Theoretical Modeling}
\label{sec:modelling}

Self-consistent modeling of jet launching is a challenging
task, especially during the earliest phases of star formation, when
the core collapse has to be modeled at the same time.
The most practical approach to study the
self-consistent jet launching during the collapse of self-gravitating
gas is through direct numerical simulation. Even then, the large
dynamical range of length scales (from the $10^4 \, \AU$ molecular
core to the sub $\AU$ protostellar disk) and time scales (from the
initial free fall time of $10^5 \, \ys$ to the orbital time of one
year or less) require expensive adaptive mesh refinement (AMR) or SPH
(smoothed particle hydrodynamics) simulations that include
magnetic fields. Furthermore, non-ideal MHD effects such as Ohmic
dissipation and ambipolar diffusion complicate the
calculations (see \S~\ref{DiskTheory}).

One of the first collapse simulations in which outflows
are observed was done more than a decade ago by \citet{Tomisaka98} with an
axisymmetric nested grid technique. These simulations of
magnetized, rotating, cylindrical cloud cores showed that a strong
toroidal field component builds up, which eventually drives a
bipolar outflow. Subsequently, a number of collapse simulations from
different groups and different levels of sophistication were
performed (among these are work by \citealt{Tomisaka02, Boss02,
  Allen03b, Matsumoto04, Hosking04, Machida04,
  Machida05a, Machida05b, Ziegler05, Machida06, Banerjee06,
  Banerjee07a, Machida07, Price07}, \linebreak
  \citealt{Machida08, Hennebelle08}, \linebreak \citealt{Duffin09,
  Mellon09, Commercon10, Burzle11, Seifried12}).
Despite the diversity in numerical approach
(e.g.\ AMR vs.\ SPH simulations) and initial problem setup, all simulations
enforce the same general picture, that magnetically launched
outflows are a natural outcome of magnetized core collapse.
The details of the outflows generated depend, of
course, on the initial parameters such as the degree of core
magnetization and the core rotation rate.

\subsubsection{Outflow Driving}

Traditionally, there is a clear distinction between outflows driven by
centrifugal acceleration \citep{Blandford82,Pudritz86,Pelletier92}
or the magnetic pressure gradient \citep{LyndenBell96, LyndenBell03}.
A frequently used quantity to make the distinction is the ratio of
the toroidal to poloidal magnetic
field, $B_{\phi}/B_{\sm{pol}}$ \citep[e.g.][]{Hennebelle08}. If this
ratio is significantly above 1, the outflow is often believed to be
driven by the magnetic pressure. However, the consideration of
$B_{\phi}/B_{\sm{pol}}$ alone can be misleading as in centrifugally
driven flows this value can be as high as 10 \citep{Blandford82}.
Close to the disk surface one can check the inclination of the
magnetic field lines with respect to the vertical axis. The field lines have to
be inclined by more than 30$^{\circ}$ for centrifugal acceleration to
work \citep{Blandford82}. Although this criterion is an exact solution
of the ideal, stationary and axisymmetric MHD equations for an outflow
from a Keplerian disk, its applicability is limited to the surface of
the disk. A criterion to determine the driving mechanism above the
disk was used by \citet{Tomisaka02} comparing the centrifugal force
$F_{\sm{c}}$ and the magnetic force $F_{\sm{mag}}$. By projecting both
forces on the poloidal magnetic field lines it can be determined which
force dominates the acceleration. For the outflow to be driven
centrifugally, $F_{\sm{c}}$ has to be larger than $F_{\sm{mag}}$.
However, for this criterion to be self-consistent the gravitational
force and the fact that any toroidal magnetic field would reduce the
effect of $F_{\sm{c}}$ have to be taken into account.

In \citet{Seifried12}, a general criterion was derived to identify
centrifugally driven regions of the outflows and to differentiate
those from magnetic pressure driven outflows. The derivation assumed
a stationary axisymmetric flow, which leads to a set
of constraint equations based on conservation laws along magnetic field
lines \citep[see also][]{Blandford82, Pudritz86, Pelletier92}. This
criterion is applicable throughout the entire outflow.

\begin{figure*}
 \centering
 \includegraphics[bb= 14 12 583 546, scale=0.3]{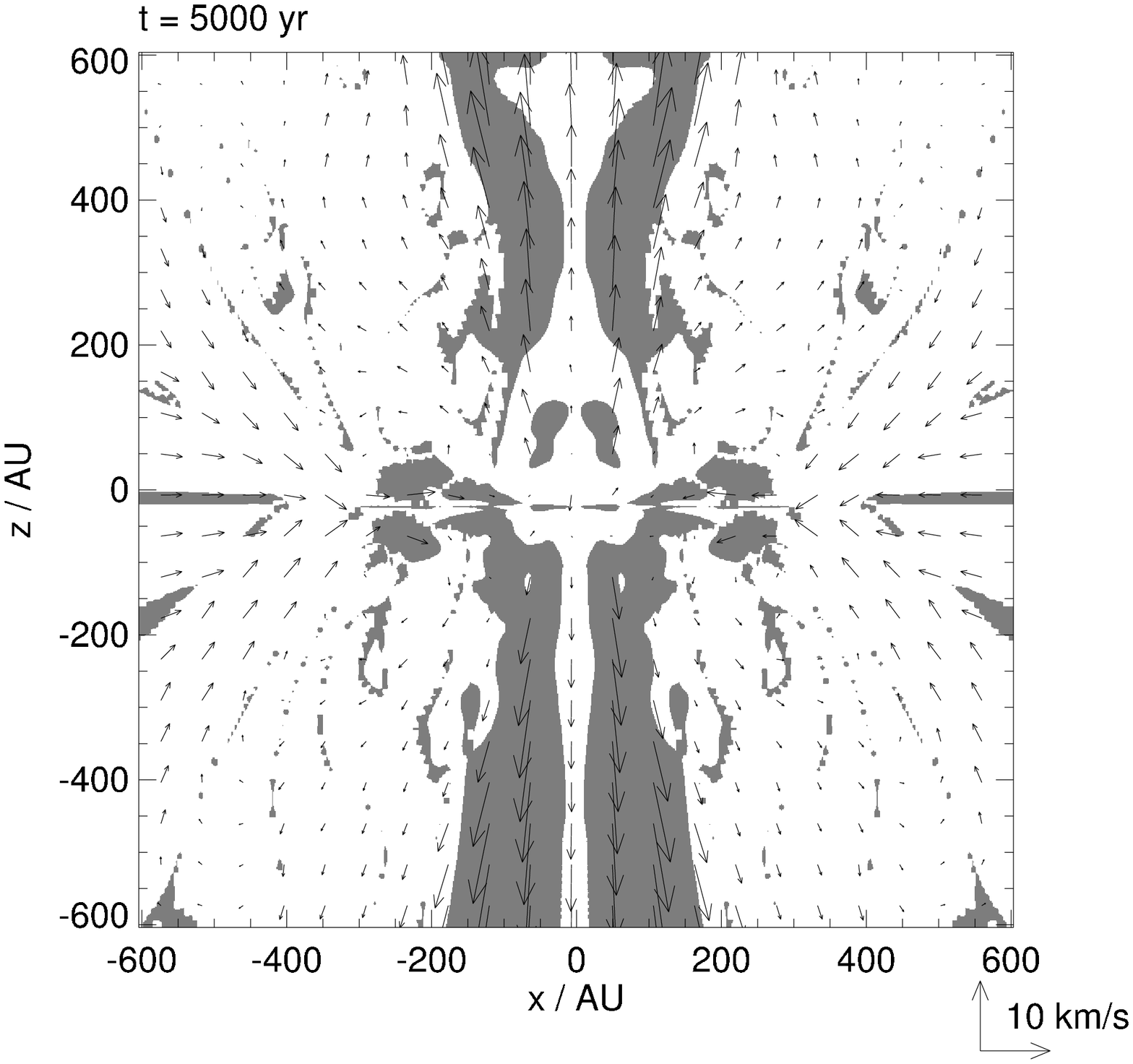}
 \includegraphics[bb= 14 12 583 546, scale=0.3]{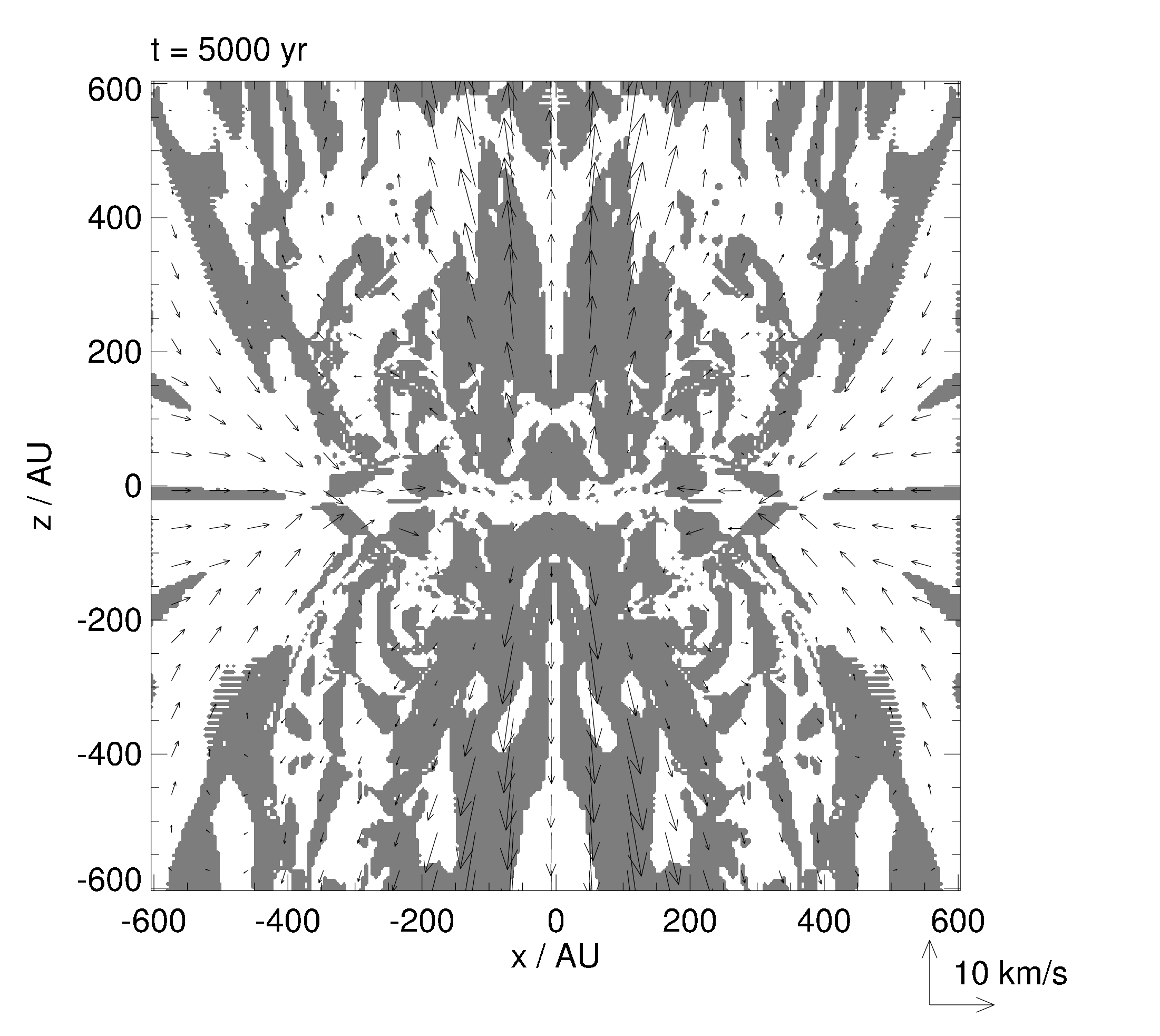} \\
 \includegraphics[bb=  4 12 583 546, scale=0.3]{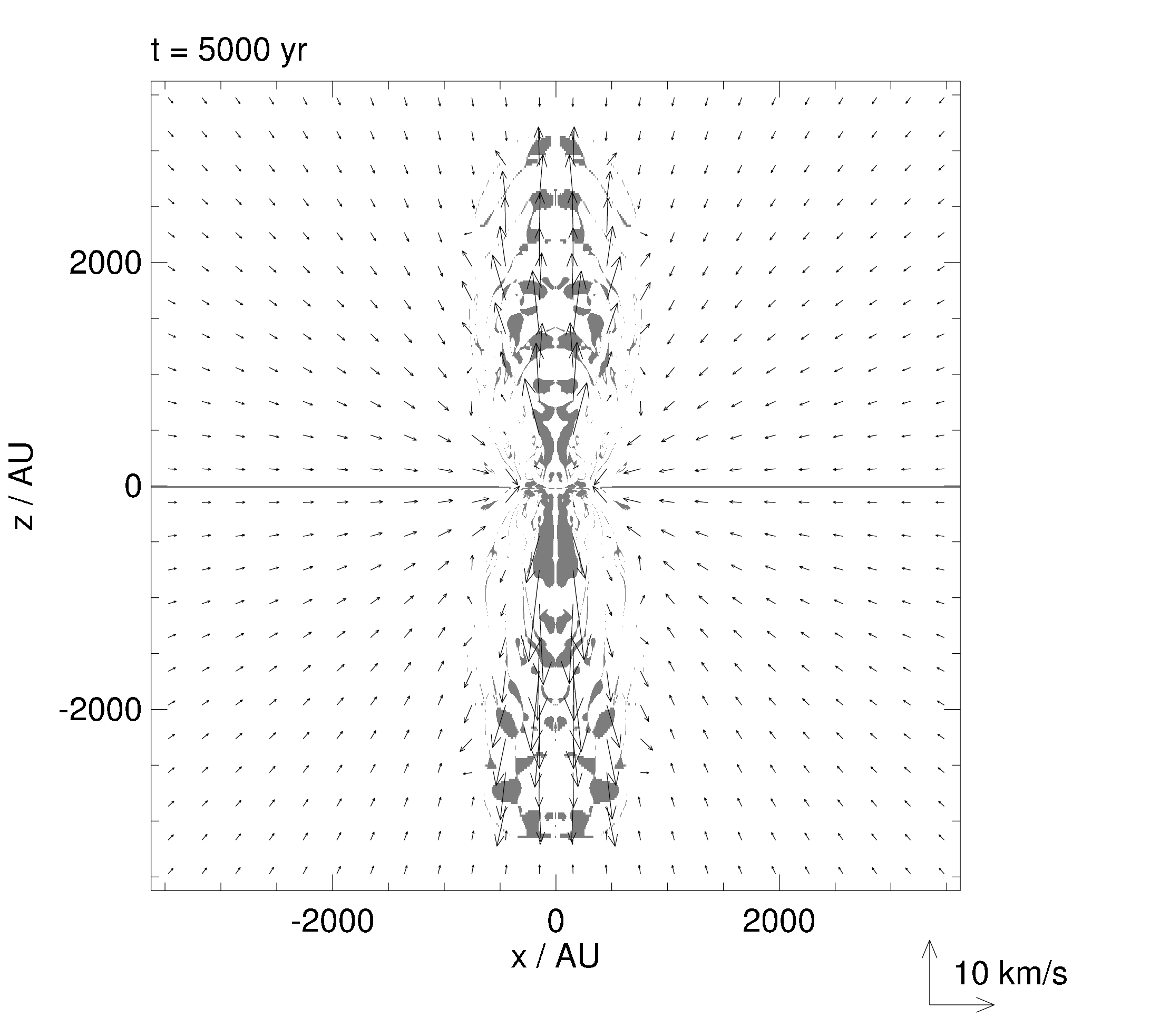}
 \includegraphics[bb=  4 12 583 546, scale=0.3]{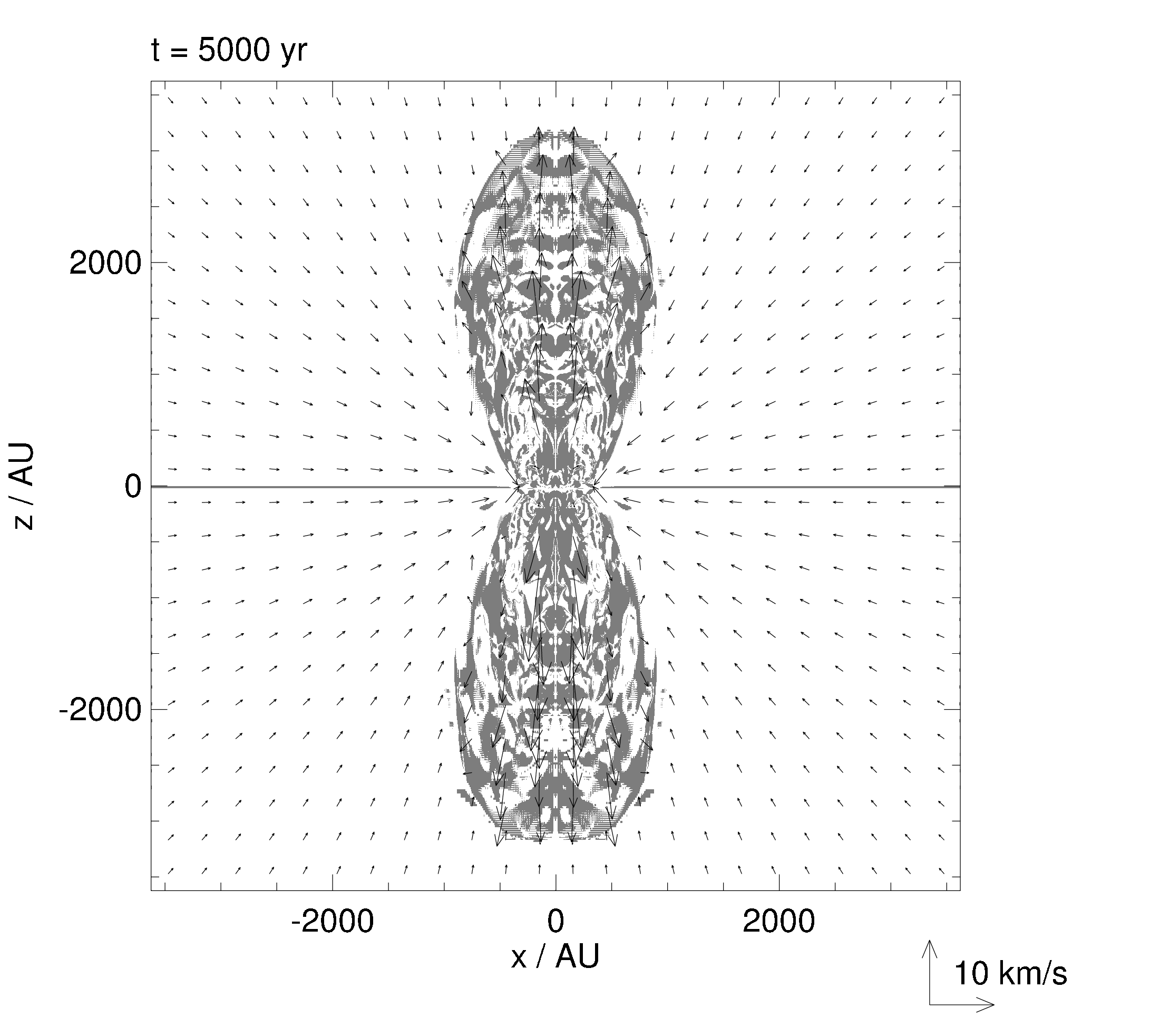}
\caption{Application of the outflow criteria derived in
  \citet{Seifried12}. The left panels show that centrifugal
  acceleration works mainly close to the $z$-axis up to a height of
  about 800 AU which agrees very well with the region where the
  highest velocities are found (see Eq.(\ref{eq:crit}). The general
  criterion (see Eq.\ \ref{eq:crit2}) is more volume
  filling and traces also regions in the outer parts [from
    \citet{Seifried12}].}
 \label{fig:BP82_26}
\end{figure*}

The general condition for outflow acceleration is \citep{Seifried12}
\begin{equation}
 \partial_{\sm{pol}} \left ( \frac{1}{2} v_{\phi}^2 + \Phi - \frac{v_{\phi}}{v_{\sm{pol}}}\frac{1}{4 \pi}\frac{B_{\phi} B_{\sm{pol}}}{\rho} + \frac{1}{4 \pi}\frac{B_{\phi}^2}{\rho} \right ) < 0
 \label{eq:crit2}
\end{equation}
where $\partial_{\sm{pol}}$ denotes the derivative along the poloidal
magnetic field.
It describes all regions of gas acceleration
including those dominated by the effect of $B_{\phi}$.  This general
outflow criterion should be compared to the case of centrifugal
acceleration where $B_\phi = 0$, i.e.\ in the case with no resulting
Lorentz force along the poloidal field line
\begin{equation}
  \frac{r}{z} \frac{1}{G M} \left(\frac{v_{\phi}^2}{r^2}(r^2 +
    z^2)^{3/2} - G M \right) \left/ \left( \frac{B_{z}}{B_{\sm{r}}} \right) \right. > 1 \; ,
 \label{eq:crit}
\end{equation}
where $M$, $r$ and $z$ are the mass of the central object, the
cylindrical radius and distance along $z$-axis, respectively.
Using both equations, one can distinguish between regions
dominated by centrifugal acceleration and those by the toroidal
magnetic pressure. Note that Eq.~(\ref{eq:crit}) does not assume a
Keplerian disk, hence it is also applicable to early-type
sub-Keplerian configurations.

An example of those outflow criteria is shown in
Fig.~\ref{fig:BP82_26} where one can see that the centrifugally
launched region is narrower and closer to the rotation axis but faster
than the outer part of the outflow. Generally, such early type
outflows are driven by both mechanisms, i.e.\ by magnetic pressure and
magneto-centrifugal forces, but the centrifugal launching should
become more dominant while the underlying disk evolves towards a more
stable Keplerian configuration.

Another, indirect, support for the outflow generation mechanism
involving magnetic driving comes from a recent
numerical study by \citet{Peters12a}. Their simulations,
which include feedback from ionizing radiation from massive
protostars, show pressure driven bipolar outflows reminiscent of
those observed around massive stars. But detailed analysis
through synthetic CO maps show that the pressure driven outflows
are typically too weak to explain the observed ones \citep[e.g.,
outflows in G5.89][]{Puga06, Su12}. The failure suggests that
a mechanism other than ionizing radiation must be found to drive
the massive outflows. Since massive star forming regions are
observed to be significantly magnetized (e.g., \citealt{girart09} and
\citealt{Tang09}), the magnetic field is a natural candidate
for outflow driving: the observed massive outflows might be driven
magnetically by the massive stars themselves, or by the collection
of lower and intermediate-mass stars in the young massive cluster.

\subsubsection{Outflow Collimation}

A general finding of self-consistent numerical simulations is that
the degree of outflow collimation is time dependent and depends on
the initial field strength. At very early stages (10$^3$ --
10$^4$ yr) outflows in typically magnetized, massive cores (with
mass-to-flux ratios of $\lambda \la 5$) are found to be poorly collimated
with collimation factors of 1 -- 2 instead of 5 -- 10, still in
agreement with observations of outflows around most young
massive protostellar
objects \citep[e.g.][]{Ridge01,Torrelles03,Wu04,Sollins04,Surcis11}.
It is suggestive that during the earliest stage, i.e.\ before
the B1--B2-type phase of the scenario described by \citet{Beuther05},
the outflows are rather poorly collimated except in case of an
unusually weak magnetic field. In their further evolution, however,
the collimation will increase quickly due to the development of a
fast, central jet coupled to the build-up of a Keplerian disk.
Therefore it might be problematic to directly link the evolutionary stage of the
massive young stellar object to the collimation of the observed
outflow as suggested by \citet{Beuther05}. Additional difficulties to
correlate ages of YSOs and the collimation of outflows arise from the
fragmentation of massive disks.
For instance, circum-system outflows (from around binaries or higher
multiples) from large sub-Keplerian disks at early stages are
possible. But those outflows are often uncollimated and might even
show spherical morphologies due to fragmentation of the highly unstable
accretion disk \citep[e.g.,][]{Peters11a, Seifried12}. Although, it
seems likely that the subsequent outflows from around single massive
protostars should be collimated, the evidence from numerical
simulations of clustered star formation showing the self-consistent
launching of such outflows
is still missing due to the lack of resolution.

Direct confirmation of those evolutionary scenarios is difficult as
one would need independent information of the age of the YSOs and
details of the magnetization of the environment \citep[see also the
discussion of][]{Ray09}. Such observations are hard to obtain and
therefore rather rare. However, there is an interesting observation
that supports the picture of very early stage, poorly collimated
outflows successively collimating over time. Observing two spatially
adjacent, massive protostars in the star forming region W75N,
\citet{Torrelles03} and \citet{Surcis11} find the younger of the two
having a spherical outflow whereas the more evolved protostar has a
well collimated outflow. Due to their close proximity to each other,
they should have similar environmental conditions. Therefore the
difference should rather be a consequence of different evolutionary
stages, where the younger, poorly collimated outflow is possibly only
a transient.

\subsection{Feedback by Jets and Outflows}

As mentioned earlier, jets and outflows are already present at very
early stages of star formation. Hence, their influence on the subsequent
evolution within star forming regions may not be neglected.  In
particular, in cluster forming regions, outflows are believed to
influence or even regulate star formation (see also chapters by
Krumholz et al.\ and Frank et al.\ in this volume) as originally
proposed by \citet{Norman80}. Since
then a number of numerical simulations tried to address this issue
\citep[e.g.,][]{Li06, Banerjee07b, Nakamura07, Banerjee09, Wang10,
  LiZY10, Hansen12}, but with different outcomes. Detailed single jet
simulations demonstrated that the jet power does not couple
efficiently to the ambient medium and is not able to drive
volume-filling supersonic turbulence \citep{Banerjee07b}. This is
because the bow shock of a highly collimated jet and developed
jet instabilities mainly excite sub-sonic velocity fluctuations.
Similarly, the simulations by \citet{Hansen12} showed that
protostellar outflows do not significantly affect the overall cloud
dynamics, at least in the absence of magnetic fields. Otherwise, the
results from simulations of star cluster forming regions
\citep[e.g.,][]{Li06, Nakamura07, Wang10, LiZY10} clearly show an
impact of outflows on the cloud dynamics, the accretion rates and the
star formation efficiency. But this seems to be only effective if
rather strong magnetic fields and a high amount of initial turbulence
are present in those cloud cores. The jet energy and momentum are
better coupled to a turbulent ambient medium than a laminar one
\citep{Cunningham09}.

\subsection{Future Research on Outflows around Protostars}

Undoubtedly, jets and outflows from YSOs are strongly linked through
the magnetic field to both the disk and surrounding envelope.
Deciphering the strength and morphological structure of magnetic
fields of jets and outflows will be a key to gain a better
understanding of these exciting phenomena. Unfortunately, there are very few
direct measurements of magnetic field strengths in YSO jets to date.
The situation should improve with the advent of new radio
instruments \citep[see also][]{Ray09}. For example, the
{\it Magnetism Key Science Project} of LOFAR ({\tt http://www.lofar.org})
plans to spatially resolve the polarized structure of protostellar
jets to examine their magnetic field structure and to investigate
the impact of the field on the launching and evolution of protostellar
jets. Prime targets would be
the star forming regions of Taurus, Perseus \& Cepheus Flare
molecular clouds with sub-arc-second resolution. The
forthcoming {\it Square Kilometer Array}
(SKA; {\tt http://www.skatelescope.org}) will also offer unprecedented
sensitivity to probe the small scale structure of outflows around
protostars \citep{gSKA-WP13}.

\section{Connecting Early Disks to Planet Formation }
\label{Disk2Planet}

As we have seen, the first $10^5$ years in the life of a protostellar disk are
witness to the accretion of the bulk of the disk mass, the rapid evolution of
its basic dynamics, as well as the most vigorous phase of its outflow activity.
This is also the period when the basic foundations of
the star's planetary system are laid down.  Giant planet formation
starts either by rapid gravitationally driven fragmentation
in the more distant regions of massive disks
\citep[e.g.,][]{Mayer02, Rafikov09},
or by the formation of rocky planetary cores that over
longer (Myr) time scales will accrete massive gaseous envelopes
(e.g., \citealt{Pollack96}, chapter by Helled et al.).
Terrestrial planet formation is believed to occur as a consequence of the
oligarchic collisional phase that is excited by perturbations caused
by the appearance of the giant planets (see chapter by Raymond et al.).
There are several important connections between the first phases of
planet formation and the properties of early protostellar disks.  We
first give an overview of some of the essential points, before
focusing on two key issues.

At the most fundamental level, the disk mass is central to the character
of both star and planet formation. Most of a star's mass is accreted
through its disk, while at the same time, giant planets must compete
for gas from the same gas reservoir. Sufficiently massive
disks, roughly a tenth of the stellar mass, can in these early stages
generate strong spiral waves which drive rapid accretion onto the
central star.  Such disks are also prone to fragmentation. Early disk
masses exceeding 0.01 $M_{\odot}$ provide a sufficient
gas supply to quickly form Jovian planets \citep{Weidenschilling77}.
The lifetime of protostellar disks, known to be
in the range 3--10 Myr, provides another of the most demanding constraints on
massive planet formation.
Detailed studies using the Spitzer Space Telescope indicate that 80 \%
of gas disks around stars of less than 2 $M_{\odot}$ have dissipated
by 5 Myr after their formation \citep{Carpenter06,Hernandez08,Hernandez10}.

As already discussed, protostellar outflows are one of the earliest
manifestations of star formation. Class 0 sources are defined as
having vigorous outflows and this implies that magnetized disks
are present at the earliest times. Magnetic fields that thread
such disks are required for the outflow launching. These fields also
have a strong quenching effect on the fragmentation of disks which
has consequences for the gravitational instability picture of
planet formation.  How early does Keplerian behavior set in?
Some simulations \citep[e.g.,][]{Seifried12b} suggest that, even
within the first few $10^4$ years, the disk has already
become Keplerian (\S~\ref{DiskTheory}) making the launch of
centrifugally driven jets all the more efficient. Before this, it
is possible that angular momentum transport by spiral waves is significant.

Stars form as members of star clusters and this may have an effect on
disk properties, and therefore, upon aspects of planet formation.
Observations show that as much as 90 \% of the stars in the galactic disk
originated in embedded young clusters  \citep{Gutermuth09}.
The disk fraction of young stars in clusters such as $\lambda$ Orionis
\citep{Hernandez10} and other clusters such as Upper Scorpius
\citep{Carpenter06} is similar to that of
more dispersed groups indicating that the dissipative time scale
for disks is not strongly affected by how clustered the star formation
process is.  We note in passing that the late time
dissipation of disks is controlled both by photo-evaporation, which
is dominated by the FUV and X-ray radiation fields of the whole
cluster and not of the host star, as well as by the clearing of holes
in disks by multiple giant planets (see chapter by Espaillat et al.).
Calculations show that FUV radiation fields, produced mainly by
massive stars, would inhibit giant planet formation in 1/3  to 2/3
of planetary systems, depending on the dust attenuation. However,
this photo-evaporation affects mainly the outer regions of disks,
leaving radii out to 35 AU relatively unscathed \citep{Holden11}.

Rocky planetary cores with of order 10 Earth masses are essential for
the core accretion picture of giant planets.  This process
must take place during the early disk phase --- the first $10^5$ years
or so (see chapter by Johansen et al.\@) in order to allow enough time
for the accretion of a gas envelope. Therefore the appearance of
planetesimals out of which such giant cores are built, must also be
quite rapid and is another important part of the first phases of
giant planet formation in early protostellar disks.  It is important
to realize therefore,  that the various aspects of non-ideal MHD
discussed in the context of disk formation in
section~\ref{DiskTheory}, being dependent on
grain properties,  take place in a rapidly evolving situation
wherein larger grains settle, agglomerate, and go on to pebble formation.

Finally, a major factor in the development of planetary systems in
early disks  arises from the rapid migration of the forming
planetary cores.   As is well known, the efficient exchange of planetary
orbital angular momentum with a gaseous disk by means of Lindblad
and co-rotation resonances leads to very rapid inward migration of
small cores on $10^5$ year time scales (\citealt{Ida08};
see chapter by Benz et al.).   One way of drastically slowing such
migration is by means of planet traps --- which are regions
of zero net torque on the planet that occur at disk
\citep{Masset06, Matsumura09, Hasegawa11, Hasegawa12}.
Inhomogeneities in disks,  such as dead
zones, ice lines, and heat transitions regions (from viscous disk heating
to stellar irradiation domination)  form special narrow zones where
growing planets can be trapped.  The early appearance
of disk inhomogeneities and such planet traps encodes the basic
initial architecture of forming planetary systems.
We now turn to a couple of these major issues in more detail.

\subsection{Fragmentation in Early Massive Disks}

Early disks are highly time-dependent, with infall of the core
continuously delivering mass and angular momentum to the forming disk.   Moreover, the central
star is still forming by rapid accretion of material through the
disk.  Depending on the infall rate, the disk may or may not be
self-gravitating (see below).
Disk properties and masses can be measured
towards the end of this accretion phase in the Class I sources.
A CARMA (Combined Array for Research in Millimeter-wave Astronomy) survey of 10 Class I disks carried out by \citet{Eisner12}
as an example showed that only a few Class I disks exceed
$0.1$ M$_{\odot}$, and the range of masses from $<0.01$ to $>0.1$
M$_{\odot}$ exceeds that of disks in the Class II phase.  This is
a tight constraint on the formation of massive planets and already
suggests that the process must have been well on its way before
even the Class I state has been reached.

\begin{figure}[ht]
\resizebox{\hsize}{!}{\includegraphics{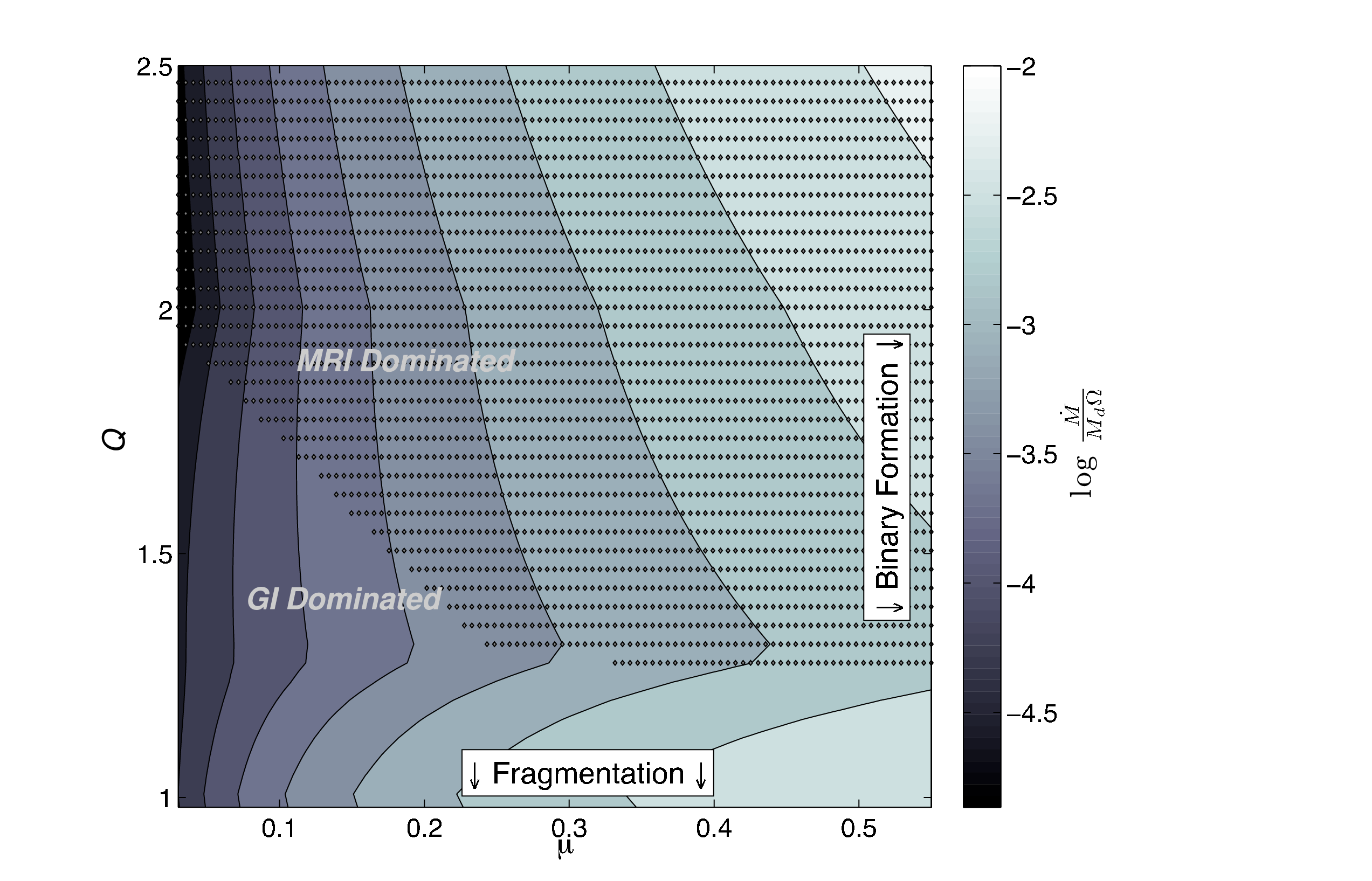}}
\caption{Contours of the dimensionless accretion rate
$\dot{M}_{*}/ ( M_{d}\Omega )$  from the disk onto the star from both
transport components of the model. The lowest contour level
is $10^{-4.8}$, and subsequent contours increase by 0.3 dex.
The effect of each transport mechanism is apparent in the curvature of
the contours.  The MRI
causes a mild kink in the contours across the Q=2 boundary and is
more dominant at higher disk masses due to the assumption of a
constant disk turbulence parameter.  Adapted from \citet{Kratter08}
}\label{fig:Kratter}
\end{figure}

Two quantities that control the gravitational stability of a
hydrodynamic disk are the ratio of the disk mass to the total mass
of the system $\mu = M_d / ( M_d + M_*) $ and the Toomre instability
parameter $Q = c_s \kappa / \pi G  \Sigma_d$, where
$\Sigma_d = M_d / (2 \pi R_d^2)$ is the surface mass density of
the disk.  Heating and cooling of the disk as well as its
general  evolution alter Q and infall onto the disk changes $\mu$
\citep{Kratter08}.  The mass balance in an early disk will
depend upon the efficiency of angular momentum transport, which
is widely believed to be governed
by two mechanisms: gravitational torque as well as the magneto-rotational instability (MRI).
Angular momentum
transport through the disk by these
agents can be treated as ``effective'' viscosities:
$\alpha_{GI} $ and $\alpha_{MRI} \simeq 10^{-2} $.  These drive the total
disk viscosity $\alpha = \alpha_{GI}(Q,\mu) + \alpha_{MRI}$.
The data produced by numerical simulations can be used to estimate
$\alpha_{GI}$ in terms of Q and $\mu$ (e.g.\ simulations of
\citealt{Vorobyov05, Vorobyov06}).
The  resulting accretion rate
onto the central star can be written in dimensionless form as ${\dot M}_* / M_d
\Omega$, where the epicyclic frequency has been replaced by the
angular frequency $\Omega$ of the disk.
The accompanying Fig.\ \ref{fig:Kratter}
shows the accretion rate through the disks in this phase space.
The figure shows that the greatest part of this disk Q-$\mu$ phase space is
dominated by MRI transport rather than by gravitational torques from
spiral waves.  Evolutionary tracks for accreting stars can be computed in this diagram, which
can be used to trace the evolution of the disks.
Low mass stars will have lower values of $\mu$,  MRI dominated
evolution,  masses of order 30 \% of
the system mass, and have typical outer radii of order 50 AU\@.
High mass stars by comparison are predicted to have high values of
$ \mu \simeq 0.35$ and
an extended period of local fragmentation as the accretion rates
peak, as well as a disk outer edge at 200 AU\@.

The fragmentation of disks is markedly affected by the presence
of significant magnetic fields. One of the main effects of a
field is to modify the Toomre criterion.  Because part of the action
of the threading field in a disk is to contribute a supportive
magnetic pressure, the Toomre $Q$ parameter is modified with the
Alfv{\'e}n velocity  $v_A$ (the typical propagation speed of a transverse
wave in a magnetic field);  $Q_M =  (c_s^2 + v_A^2)^{1/2} \kappa/\pi
G \Sigma_d $.  For typical values of the  mass to flux ratio,
magnetic energy densities in disk are comparable to thermal or
turbulent energy densities. \citet{Seifried11} find that the
magnetic suppression of disk fragmentation occurs in most of
their models. Even the presence of ambipolar diffusion of the disk
field does not significantly enhance the prospects for fragmentation
\citep{Duffin09}.

Gravitational fragmentation into planets or low mass companions
requires that
$Q \sim 1$,  however this is
not sufficient.
Fragments must also cool sufficiently rapidly as was first derived by \citet{Gammie01}, and
generalized by \citet{Kratter11}.  The condition for sufficiently
rapid cooling depends, in turn, upon how the disk is heated.
The inner regions of disks are dominated by
viscous heating, which changes into dominant radiative heating
by irradiation of the disk by the central star, in the outer regions \citep{Chiang97}.  The transition
zone between these two regions, which has been called a ``heat
transition radius'' \citep{Hasegawa11} is of importance
both from the view of gravitation fragmentation \citep{Kratter11} as
well as for the theory of planet traps. This radius occurs where
the heating of the surface of the disk by irradiation by the central star
balances the heating of the disk at the midplane by viscous heating.
Stellar irradiation dominates viscous heating if the temperature
$T$ exceeds a critical value:  $ T > [(9/8) ( \alpha \Sigma /
\sigma) ( k / \mu)  \tau_R \Omega]^{1/3},$ where $\tau_R$ is
the Rosseland mean opacity,  $\alpha$ is the viscosity parameter,
$\sigma$ is the Stefan-Boltzmann constant, $k$ is Boltzmann's constant, and
$ \Omega$ is the orbital angular frequency.

The long term survival of fragments depends upon
 three different forces including gas pressure, shearing in
the disk, and mutual interaction and collisions \citep{Kratter11}.
The role of pressure in turn
depends upon how quickly the gas can cool and upon its primary
source of energy.
Fragmentation in the viscously heated
regime --- which is where giant planets may typically form ---
can occur if the ratio of the cooling time to the dynamical time
$\beta$ is sufficiently small for a gas with adiabatic
index $\gamma$:  $\beta < ( [4/9\gamma (\gamma - 1)]
\alpha_{sat}^{-1} $, where the saturated value of the viscous
$\alpha$ parameter refers to the turbulent amplitude that can be
driven by gravitational instability.  Infall plays an important role in
controlling this fragmentation. Generally, a higher infall rate
$\dot M$ drives the value of $Q$ downwards as seen in the Figure.
Rapid infall will tend to drive disks closer to instability therefore,
and perhaps even on to fragmentation.  Irradiated disks, by
contrast, have a harder time to fragment.  The basic
point here is that while gravitationally driven turbulence can
be dissipated to maintain $Q = 1$, irradiated disks do not have
this property.  Once an irradiated disk moves into a critical Q regime,
they are more liable to fragment since there is no intrinsic
self-regulatory mechanism for maintaining disk temperature near
a critical value.   The results indicate that irradiated disks
can be driven by infall to fragment at lower accretion rates onto
the disk.

Spiral arms compress the disk gas and are the most likely sites
for fragmentation.  A key question is what are the typical masses
of surviving fragments. Recent numerical simulations of
self-gravitating disks, without infall, have gone much farther
into the non-linear regimes to follow the fragmentation into
planet scale objects \citep[e.g.,][]{Boley10,Rogers12}.
Using realistic cooling functions for disks, \citet{Rogers12} have
simulated self-gravitating disks and have found a new criterion
for the formation of bound fragments. Consider a patch of a disk that
has been compressed into a spiral arm of thickness
$l_1$ (see Figure~9).
 This arm will form gravitationally bound fragments if $l_1$ lies within the Hill radius
$H_{Hill} = [G \Sigma_l l_1^2 / 3 \Omega^2]^{1/3}$ of the arm, or:
 $ (l_1 / 2 H_{Hill}) < 1$.

\begin{figure}[ht]
\resizebox{\hsize}{!}{\includegraphics{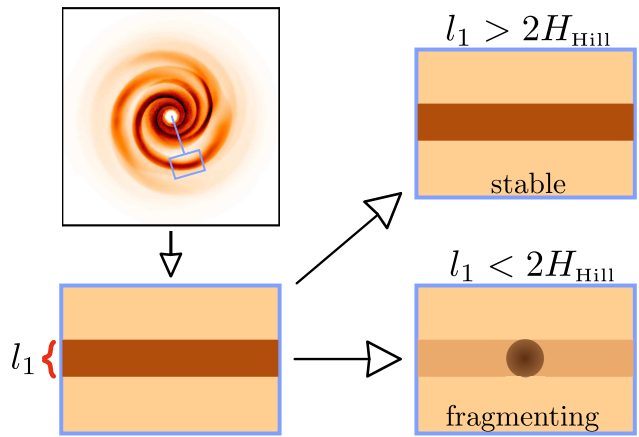}}
\caption{The Hill criterion for spiral arm fragmentation: if a piece
of the spiral arm of width $l_1$ lies within its own Hill thickness, then that
section of the arm is free to collapse and fragmentation takes place.
If such a section of the spiral arm lies outside of its own Hill
thickness, then shear stabilizes the arm and fragmentation does not
take place. Adapted from  \citet{Rogers12}
}\label{fig:RogersWadsley12}
\end{figure}

This criterion also addresses the ability of shear to prevent the
fragmentation of the arm. Numerical simulations verify that this criterion
describes the survival of gravitationally induced fragments in the
spiral arms.  The results have been applied to disks around
A stars and show that fragments of masses $15 M_{\mathrm Jup}$ can form and
survive at large distances of the order of
95 AU from the central star, in this radiation heating dominated
regime.  Brown dwarf scale masses seem to be preferred.

Finally we note that clumps formed in the outer parts of disks may
collapse very efficiently, perhaps on as little as thousand year
time scales.  This suggests that clumps leading to the formation
of giant planets could collapse quickly and survive transport to
the interior regions of the disk \citep{Galvagni12}.
We turn now to planetary transport through disks.

\subsection{Planet Traps and the Growth and Radial Migration of Planets}

The survival of planetary cores in early disks faces another classic
problem arising from the exchange of orbital angular momentum
between the low mass planetary core and its surrounding gaseous disk.
In these early phases, protoplanetary cores can raise significant wakes
or spiral waves to their interior and exterior radial regions of
the disk.  These waves in turn exert torques back on the planet
resulting in its migration.  For homogeneous disks with smooth,
decreasing density and temperature profiles, the inner wake (which
transfer angular momentum to the protoplanet --- driving it outwards)
is slightly overcome by the outer wake (which extracts orbital
angular momentum from the protoplanet driving it inwards).  This
results in a net torque which results in the rapid inward (Type I)
migration of the planet.  As is well known, Monte Carlo population
synthesis calculations on protoplanets in evolving accreting disks
find that protoplanetary cores migrate into the center of the disk
within $10^5$ years \citep{Ida05,Ida08} --- the timescale
characterizing early disks.

Rapid Type I migration must be slowed down by at least a factor
of 10 to make it more compatible with the lifetime of the disk
(see chapter by Benz et al.).  As noted, this can be achieved in
disks at density and temperature inhomogeneities at whose
boundaries, migration can be rapidly slowed, or even stopped.
 We have already encountered two of these in other contexts,
namely dead zones and heat transitions wherein there is a change
in density and in disk temperature gradient respectively.
Heat transitions were shown to be planet traps by \citep{Hasegawa11}.
A third type of trap is the traditional
ice line, wherein the temperature transition gives rise to a change
in disk opacity with a concomitant  fluctuation in the density
\citep{Ida08}. \citet{Hasegawa11,Hasegawa12} showed that
protoplanetary cores can become trapped at these radii.

Dead zone edges act to stop planetary migration as has been shown
in theory \citep{Masset06} and numerical simulations
\citep{Matsumura09}. A region that is starved of ionization, which occurs
where disk column densities are high enough to screen out ionizing cosmic
rays, has large Ohmic dissipation which prevents the operation of
the MRI instability.  This region is known as a dead zone because
it is unable to generate the MHD turbulence necessary
to sustain a reasonable viscosity $\alpha_{MRI}$ \citep{Gammie96}.
A dead zone is most likely to be present in the inner regions of disks in their
later stages.   During the earliest stages of disk evolution,
when the disks are the most massive, these would extend out to 10 AU or so
\citep{Matsumura06}. It is necessary to  transport away
disk angular momentum to allow an ongoing large accretion
flow that is measured for young stars.   Older models
supposed that a thin surface layer is sufficiently well ionized to
support MRI turbulence.
Recent simulations show, however, that such MRI active surface layers
may not occur \citep{Bai13}. The inclusion of all three
of the non-ideal effects discussed in \S~\ref{DiskTheory} (Ohmic
resistivity, Hall effect, and ambipolar diffusion) strongly
changes the nature of MHD instability in vertically stratified disks.
Going from the dense mid plane to the surface of the disk, the
dominant dissipation mechanism changes:  from Ohmic resistivity at
the mid plane, to Hall effect at mid-scale heights, to
ambipolar diffusion in the surface layers.   The results show that ambipolar diffusion
shuts down the MRI even in the surface layers in the presence of a
net (non-zero) magnetic flux through the disk.    Instead of the
operation of an MRI in the dead zone, a strong MHD disk wind is
launched and this carries off the requisite disk angular momentum
very efficiently (e.g.\ review; \citealt{Pudritz07},
\S~\ref{outflow} of this paper).

Planetary cores accrete most of their mass while moving along with
such traps.  As the disk accretion rate falls in an evolving disk, the
traps move inward at different rates.  At early times, with high
accretion rates, the traps are widely separated.
As the disk accrete rate falls from a high of
$10^{-6}$M$_{\odot}$ yr$^{-1}$ to lower values at later
times, the traps slowly
converge at small disk radii, which likely initiates planet-planet interactions.

In summary then, this section has shown that the properties of
early magnetized disks as characterized by high infall rates
and disk masses, as well as powerful outflows, can strongly
influence the early phases of planet formation and migration.

\section{Synthesis}
\label{Synthesis}

Both observational and theoretical studies of disk formation are
poised for rapid development. Observationally, existing dust
continuum surveys of deeply embedded ``Class 0'' objects indicate
that a compact emission component apparently distinct from the
protostellar envelope is often present. Whether this component is
a rotationally supported disk (RSD) or not is currently unclear in
general. With unprecedented sensitivity and resolution, ALMA should
settle this question in the near future.

On the theoretical side, recent development has been spurred largely
by the finding that magnetic braking is so efficient as to prevent
the formation of a RSD in laminar dense cores magnetized to realistic
levels in the ideal MHD limit --- the so-called ``magnetic braking
catastrophe.'' Although how exactly this catastrophe can be avoided
remains unclear, two ingredients emerge as the leading candidates
for circumventing the excessive braking --- Ohmic dissipation and
complex flow pattern (including turbulence, misalignment of
magnetic field and rotation axis, and possibly irregular core
shapes): the former through the
decoupling of magnetic field lines from the bulk neutral matter at
high densities and the latter through, at least in part,
turbulence/misalignment-induced magnetic reconnection.
It is likely that both ingredients play a role, with Ohmic dissipation
enabling a dense, small (perhaps AU-scale) RSD to form early in
the protostellar accretion phase, and flow complexity facilitating
the growth of the disk at later times by weakening the magnetic
braking of the lower density protostellar accretion flow.

The above hybrid picture, although probably not unique, has the virtue
of being at least qualitatively consistent with the available
observational and theoretical results. The small Ohmic
dissipation-enabled disk can in principle drive powerful outflows
that are a defining characteristic of Class 0 sources from close
to the central object where the magnetic field and matter are well
coupled due to thermal ionization of alkali metals. During the Class
0 phase, it may have grown sufficiently in mass to account for the
compact component often detected in interferometric continuum
observations, but not so much in size as to violate the constraint
that the majority of the compact components remain unresolved to
date. The relatively small size of early disks could result from
magnetic braking.

A relatively small early disk could also result from a small
specific angular momentum of the core material to begin with.
There is a strong need to determine more systematically the
magnitude and distribution of angular momentum in prestellar
cores through detailed observations. Another need is to
determine the structure of the magnetic fields on the
$10^2$--$10^3$~AU scale that is crucial to disk formation. For
example, detection of magnetic field twisting would be direct
evidence for magnetic braking. With the polarization capability
coming online soon, ALMA is expected to make progress on this
observational front.

On the theory front, there is a strong
need to carry out simulations that combine non-ideal MHD effects
with turbulence and complex initial conditions on the core scale,
including magnetic field-rotation misalignment. This will be
technically challenging to do, but is required to firm up disk
formation scenarios such as the hybrid one outlined above.

All of the evidence and theory shows that the formation of outflows is
deeply connected with the birth of magnetized disks during
gravitational collapse.  Early outflows and later higher speed jets
may be two aspects of a common underlying physical picture in which
acceleration is promoted both by toroidal magnetic field pressure on
larger scales as well as centrifugal ``fling'' from smaller scales.
Simulations, theory, and observations are converging on the idea that
the collapse and outflow phenomenon is universal covering the full
range of stellar mass scales from brown dwarfs to massive stars.
Finally, the earliest stages of planet formation take place on the
very same time scales as disks are formed and outflows are first
launched.  While little is yet known about this connection, it is
evident that this must be important.  Many aspects of planet formation
are tied to the properties of early disks.

In summary, firm
knowledge of disk formation will provide a solid
foundation for understanding the links between early disks, outflows, and
planets, opening the way to discovering the deep connections
between star and planet formation.

\textbf{ Acknowledgments.}
ZYL is supported in part by NASA NNX10AH30G, NNX14AB38G,
and NSF AST1313083, RB by the Deutsche
Forschungsgemeinschaft (DFG) via the grant BA 3706/1-1, REP by a
Discovery Grant from the National Science and Engineering Research
Council (NSERC) of Canada, and JKJ by a Lundbeck Foundation Junior
Group Leader Fellowship as well as Centre for Star and Planet
Formation, funded by the Danish National Research Foundation and
the University of Copenhagen's programme of excellence.

\end{document}